\documentclass[12pt]{article}
\usepackage[a4paper]{geometry}

\usepackage{amssymb}
\usepackage{mathrsfs}
\usepackage{amsthm}
\usepackage{amsmath}
\usepackage{dsfont}
\usepackage{textcomp} 
\usepackage{physics}
\usepackage{enumitem}

\usepackage[T1]{fontenc}	

\usepackage[english]{babel}
\usepackage{graphicx}
\usepackage{tikz}
\usepackage{pgfplots}
\usepackage{hyperref}
\usepackage{url}

\newtheorem{thm}{Theorem}

\newtheorem{definition}[thm]{Definition}
\newtheorem{example}[thm]{Example}

\theoremstyle{remark}

\title{Groupoid and algebra of the infinite quantum spin chain}

\author{Florio~Maria~Ciaglia$^{1,7}$, Fabio~Di~Cosmo$^{1,2,8}$, Paolo~Facchi$^{3,4,9}$,\\ Alberto~Ibort$^{1,2,10}$, Arturo~Konderak$^{3,4,11}$, Giuseppe~Marmo$^{5,6,12}$\\
	\footnotesize{$^{1}$\textit{Depto. de Matem\'aticas, Univ. Carlos III de Madrid, Legan\'es, E-28911, Madrid, Spain}} \\
	\footnotesize{$^{2}$\textit{ ICMAT, Instituto de Ciencias Matem\'{a}ticas (CSIC-UAM-UC3M-UCM)}} \\
	\footnotesize{$^{3}$\textit{Dipartimento di Fisica and MECENAS, Università~degli~studi~di~Bari, Bari, I-70126, Italy}}\\
	\footnotesize{$^{4}$\textit{INFN, sezione di Bari, Bari, I-70126, Italy}}\\
	\footnotesize{$^{5}$\textit{ Dipartimento di Fisica ``E. Pancini'', Universit\`a di Napoli Federico II,  Naples, Italy}} \\
	\footnotesize{$^{6}$\textit{ INFN-Sezione di Napoli, Naples, Italy}} \\
	\footnotesize{$^{7}$\textit{ e-mail: \texttt{fciaglia@math.uc3m.es
	}}} \,\, 
	\footnotesize{$^{8}$\textit{ e-mail: \texttt{fcosmo@math.uc3m.es}}} \\
	\footnotesize{$^{9}$\textit{ e-mail: \texttt{paolo.facchi@ba.infn.it}}} \,\,  
	\footnotesize{$^{10}$\textit{ e-mail: \texttt{albertoi@math.uc3m.es}}} \\ 
	\footnotesize{$^{11}$\textit{ e-mail:\texttt{arturo.konderak@ba.infn.it}}}
	\footnotesize{$^{11}$\textit{ e-mail:\texttt{marmo@na.infn.it}}}
}
\date{}
\begin{document}
	\maketitle

\begin{abstract}
	It is well known that certain features of a quantum theory cannot be described in the standard picture on a Hilbert space. In particular, this happens when we try to formally frame a quantum field theory, or a thermodynamic system with finite density. This forces us to introduce different types of algebras, more general than the ones we usually encounter in a standard course of quantum mechanics. We show how these algebras naturally arise in the Schwinger description of the quantum mechanics of an infinite spin chain. In particular, we use the machinery of Dirac-Feynman-Schwinger (DFS) states developed in recent works to introduce a dynamics based on the modular theory by Tomita-Takesaki, and consequently we apply this approach to describe the Ising model.
\end{abstract}


\section{Introduction}
In his formulation of quantum theory, Schwinger introduced an approach based on the concept of transition~\cite{Schwinger_2018}. He identified the main feature that characterizes quantum mechanics in the transition between different results caused by a measurement. Subsequent measurements of non compatible observables can modify the results of an experiment, unlike classical theories where all the observables can --in principle-- be measured at the same time, and no transitions occur. From Schwinger's point of view, such description of quantum mechanics would need a ``new symbolic language'' more suitable to microscopic phenomenology. The fundamental cell of this language is the concept of selective measurement, i.e., ``a device that selects the system (in a given ensemble) with well defined values `$a$' for the observable $A$, and returns it with well defined values `$b$' for the observable $B$''. These symbols, as abstract entities, satisfy a set of axioms among which a composition law. Therefore, in a series of papers~\cite{gentle,groupoidI,groupoidII}, some of the authors proposed that a proper abstract frame for Schwinger's approach is the composition law of a groupoid. Then, using representation theory~\cite{ibort2019introduction}, it is possible to get the Dirac picture of standard quantum mechanics on Hilbert spaces~\cite{groupoidI}, whereas, by introducing the idea of groupoid algebras, it is also possible to obtain the algebraic picture of quantum mechanics~\cite{groupoidII,lgcqt}.

In order to clarify these ideas, consider the following simple example, which is a standard situation in quantum mechanics. Suppose that the spin along the $z$ axis is measured for a particle with spin $S=1/2$, and suppose that no transition between the $\ket\uparrow$ and $\ket\downarrow$ states are allowed. Let the state of the system be a coherent superposition $\psi=\alpha \ket\uparrow+\beta\ket\downarrow$, with $\abs{\alpha}^2+\abs{\beta}^2=1$. By measuring the spin along $z$ axis, we can only measure the probabilities $\abs{\alpha}^2$ and $\abs{\beta}^2$, which do not characterize the state $\psi$ at all. In particular, this state will behave like a classical mixture
\begin{equation}
	\rho=\abs {\alpha}^2\ketbra{\uparrow}{\uparrow} +{\abs{\beta}^2}\ketbra{\downarrow}{\downarrow},
\end{equation} 
with no quantum coherence between the operational (incoherent) basis $\{\ket \uparrow,\ket\downarrow\}$. It is clear that the state can be fully characterized only with other observables (the spin along $x$ or $y$ axes), which in terms of Schwinger's picture means transitions in the form $\ketbra{\uparrow}{\downarrow}$ and $\ketbra{\downarrow}{\uparrow}$. Quantum theory is encoded in the off-diagonal terms of the density matrix. This approach is used in the context of quantum resource theory of coherence, where the resource is identified in the coherent superposition of the incoherent basis~\cite{pleniocoherence,gramegna}.

Finite groupoids have been largely studied in previous works, and one can show that the general situation in this case can be summarized as a mixing between a quantum theory (transitions) and a classical one (superselection rules)~\cite{groupoidI,groupoidII}. In particular, from the algebraic perspective, one obtains standard quantum mechanics with a finite number of degrees of freedom described by the direct sum of complete algebras of bounded operators (type $I$ algebras).

In this paper, we are going to study from the point of view of Schwinger's picture of quantum mechanics a possible way to obtain families of more general algebras than type $I$. These algebras were introduced in~\cite{onringoperator,onringoperatorii,onringoperatoriii,onringoperatoriv} as nontrivial examples of independent quantum mechanical systems, and to extend the boolean algebra to the non-commutative case. As we will see later, they are classified as type $II$ and type $III$ algebras, depending on the finiteness of their projections. Introduced as abstract mathematical entities, it has been proved that they also play a central role in the algebraic description of quantum field theories~\cite{Araki1964,Fredenhagen1985,Haag1996,YNGVASON2005135}. From a mathematical point of view, it has been already shown that the algebras associated to special types of groupoids can be of type $II$ or $III$~\cite{TakesakiIII}. Here, we will exploit these considerations to investigate possible implications from the point of view of Schwinger's picture of quantum mechanics. 

In particular, we will consider a groupoid with a space of objects $\Omega_{\infty}$ made of infinite sequences $x=\left\lbrace x_i\mid x_i=0,1 \; \forall i \in \mathbb{N} \right\rbrace$ and transitions given by the action of a group $\Gamma \subset \Omega_{\infty}$ whose elements are sequences which differ from zero only for a finite number of elements $x_i$. From a physical point of view, since the groupoid of pairs of set with two objects is related to the qubit, the groupoid studied here could describe an infinite chain of qubits, where only the transitions that can flip a finite number of them are allowed. We will show that the algebra associated with such a system can be of either type \textit{II} or \textit{III}, depending on a certain probability $\lambda$, reproducing in this framework a well known result in the theory of von Neumann algebras~\cite{pukanski_some,glimms61,Powers1967}.

One of the ingredients that has been introduced in previous works on the groupoid approach to quantum mechanics is the concept of Dirac-Feynman-Schwinger (DFS) states, which are defined from functions that transform according to the underlying composition law of the groupoid, see section~\ref{sec:DFS_states}. In this paper we will investigate the structure of the family of the DFS states associated with the above-mentioned groupoid of sequences, and show their connection with cohomology of groups. In a previous work it has been put in evidence the dynamical role of these DFS states when considering the groupoid of paths associated with a kinematical groupoid~\cite{ciagilapropagator,ciaglialagrangian}. In that case, the exponent of the DFS function associated with the DFS state is interpreted as an action functional. In this work we are presenting a different way according to which DFS functions can define a dynamics. Indeed, the same exponent can be used to define a positive-valued homomorphism of the groupoid. Then, by properly choosing  a measure on the groupoid, this homomorphism determines a modular function whose associated modular automorphism determines a canonical evolution of the groupoid-algebra. As a final consideration we will investigate a concrete example of a DFS function which would reproduce the dynamics of an Ising chain in terms of the Tomita-Takesaki modular theory. In this sense, this work is an attempt to investigate more directly the dynamical content of the DFS states in a non-trivial example, towards a deeper understanding of the Schwinger's quantum action principle.  

Let us briefly outline the content of the paper. In Section~\ref{sec:algebras_review} we summarize some basic facts on $C^*$-algebras, von Neumannn algerbas and left-Hilbert algebras which will be used in the core of the paper. In Section~\ref{sec:groupoid_and_groupoid_algebras}, we will briefly present the construction of the (reduced) groupoid von Neumann algebra, whereas in Section~\ref{ch:infinite_qubit_chain_groupoid} we will apply this construction to a specific groupoid. This groupoid is built out of the action of a countable group acting freely, ergodically and non-transitively on a measurable space of sequences. By using well-known results in the theory of von Neumann algebras, we show that the groupoid algebra associated with this groupoid can be of type $II_1$ or type $III_{\lambda}$ depending on the choice of the measure on the space of sequences. In Section~\ref{sec:DFS_states}, finally, we provide an algorithmic construction for the DFS functions and their relations to cohomology of groups. We conclude the Section with an explicit construction of a measure on the groupoid under investigation whose modular function can be interpreted as the dynamics of an infinite Ising chain. 

\section{Algebraic description}
\label{sec:algebras_review}
Quantum mechanics has different (in general nonequivalent) descriptions. The standard one, associated with the names of Schr\"odinger, Dirac and Von Neumann, is based on the choice of a Hilbert space, whose rays represent the states of the system~\cite{dirac2013principles}. Observables enter this picture as bounded self-adjoint operators on the chosen Hilbert space.

A different description of quantum mechanics, which has roots in Heisenberg's approach to quantum mechanics and later developed by Araki, Haag, Kastler in order to axiomatically introduce the notion of quantum fields, is obtained in terms of algebras~\cite{araki1999mathematical,blackadar2006operator,bratteli2012operator}. Here, observables are the building blocks of the theory, as the self-adjoint elements of a given algebra, while states are the positive functionals over this algebra. Once a state is chosen, a Hilbert space can be built via the GNS-construction and the algebra is represented in terms of bounded self-adjoint operators, providing a connection with the previous approach.

Different types of algebras can be chosen, depending on the theory under analysis. In this chapter, we will give a concise introduction on $C^*$-algebras, $W^*$-algebras and left-Hilbert algebras, concepts which will be widely employed in the following sections. See, for example~\cite{blackadar2006operator, bratteli2012operator,TakesakiI,TakesakiII,sakai2012c} for a more detailed analysis. All of them are involutive algebras (or $*$-algebras) over the complex field $\mathbb{C}$, which means, a vector space $\mathfrak{A}$ over $\mathbb{C}$, equipped with a multiplication law which associates the element $AB$ to every pair $A, \,B$ of elements of $\mathfrak{A}$. The product is assumed to be associative ($A(BC) = (AB)C$, $\forall A,B,C \in \mathfrak{A}$), and distributive ($A(B+C) = AB + AC$, $\alpha\beta(AB) = (\alpha A)(\beta B)$, $\forall A,B,C \in \mathfrak{A}$ and $\forall \alpha, \beta \in \mathbb{C}$). In addition this algebra is endowed with a map, called involution, $*:\mathfrak A\rightarrow \mathfrak A$ satisfying 
\begin{enumerate}
	\item[i.] $(AB)^*=B^*A^*$,
	\item[ii.] $A^{**}=A$,
	\item[iii.] $(A+\lambda B)^*=A^*+\overline \lambda B^*$,
\end{enumerate}
with $A,B$ in $\mathfrak A$, and $\lambda$ in $\mathbb C$.

The differences among these type of algebras are mainly due to the way we introduce a topology on these spaces. We will point them out in the rest of this section.    

\subsection{$C^*$-algebras and $W^*$-algebras}\label{sec:C-W algebras}
The concept of $C^*$-algebra was formalized by Segal~\cite{segal}, following the work of Gelfand and Naixmark~\cite{gelfandNormierte,gelfand1943imbedding}. It is defined as a complex involutive algebra $\mathfrak A$ on which a norm $\norm{\cdot}$ is defined, which satisfies
\begin{enumerate}
	\item[i.]	$\norm{AB}\leqslant \norm{A}\norm{B}$,
	\item[ii.] $\norm{A^*A}=\norm{A}^2,$
\end{enumerate}
and $\mathfrak A$ is complete with respect to this norm. An immediate consequence of the previous properties is that $\norm{A^*}=\norm{A}$. This implies that all the operations characterizing the algebra are continuous with respect to the topology induced by this norm, making $\mathfrak{A}$ a Banach $*$-algebra. As we shall see, the difference between $C^*$-algebras and von Neumann algebras lies in the topology used to close the algebra. According to Segal, this uniform topology is more natural in physics. In particular, this structure allows us to extend some notions of operator analysis (such as spectral analysis or decomposition theory) to the algebraic framework~\cite{bratteli2012operator}.

In the algebraic description, states enter the theory as positive normalized functionals over $\mathfrak{A}$. Explicitly, they are defined as linear maps
\begin{equation}
	\omega:\mathfrak A\rightarrow \mathbb C
\end{equation}
satisfying
\begin{enumerate}
	\item[vi.] $\omega(A^*A)\geqslant0\quad A \in\mathfrak A$,
	\item[vii.] $\norm{\omega} = 1$.
\end{enumerate}

When the algebra has a unit $\mathbb I$, $\norm{\omega}=1$ is equivalent to the condition $\omega(\mathbb I)=1$. One can prove that, see~\cite{blackadar2006operator,bratteli2012operator}, any such functional is continuous in the uniform topology and satisfies a Cauchy-Schwarz type inequality. In the algebraic approach to quantum mechanics, states are the crucial ingredient for the statistical interpretation of the theory, since for a self-adjoint element, $A=A^*$ in $\mathfrak A$, $\omega(A)$ is interpreted as the expectation value of the measurement $A$. Their relation with states of an Hilbert space can be obtained from representation theory, namely, algebra homomorphisms $\pi$ from $\mathfrak A$ into the set of bounded operators over a Hilbert space $\mathcal B(\mathcal H)$.

In particular, given a state $\omega$ over a $C^*$-algebra $\mathfrak A$, there exists a unique (up to a unitary transformation) representation $(\mathcal H_{\omega}, \pi_{\omega})$, with a vector state $\Psi_\omega\in\mathcal H_\omega$ being a representative of $\omega$:
\begin{equation}\label{eq:staterepresentativegns}
	\braket{\Psi_\omega}{\pi_\omega(A)\Psi_\omega}=\omega(A),\qquad\forall A\in\mathfrak A.
\end{equation}
Moreover, $\Psi_\omega$ is cyclic, namely, $\{\pi_\omega(A)\Psi_\omega|A\in\mathfrak A\}$ is dense $\mathcal H_\omega$. This representation is usually called cyclic or GNS-representation~\cite{segal,gelfand1943imbedding}. We will indicate it with the triplet
\begin{equation}\label{eq:gnsrepresentation}
	(\mathcal H_\omega,\pi_\omega,\Psi_\omega).	
\end{equation}
Using this representation and equation~\eqref{eq:staterepresentativegns}, we see that a state can always be thought as a vector state in a proper representation. 

Von Neumann algebras, or $W^*$-algebras, originally introduced as rings of operators, are $C^*$-algebra which are the dual of a Banach space, i.e., for every $W^*$-algebra $\mathfrak{M}$ there exists a Banach space $\mathfrak{M}_{\ast}$ such that $\mathfrak{M} = \mathfrak{M}_{\ast}^*$. It was Sakai~\cite{sakai2012c} who provided an abstract approach for these algebras detached from the introduction of a Hilbert space.
Indeed, in the original formulation by von Neumann, they are defined starting from a Hilbert space $\mathcal H$: they are subalgebras of the algebra of bounded operators $\mathcal B(\mathcal H)$. Adding some flesh, given a set of operators $\mathfrak M\subset \mathcal B(\mathcal H)$, we call its commutant the set $\mathfrak M'$ of the operators in $\mathcal B(\mathcal H)$ which commute with every element in $\mathfrak M$:
\begin{equation}
	\mathfrak M'=\{A\in\mathcal B(\mathcal H)|AB=BA, \quad\forall B\in\mathfrak M\}.
\end{equation}
Instead, the bicommutant of $\mathfrak M$ is the commutant of $\mathfrak M'$:
\begin{equation}
	\mathfrak M''=(\mathfrak M')'\supset\mathfrak M.
\end{equation}
A von Neumann algebra $\mathfrak M$ on a Hilbert space $\mathcal H$ is a $*$-algebra of bounded operators for which $\mathfrak M=\mathfrak M''$. It is easy to see that a von Neumann algebra is also a $C^*$-algebra, as it is closed under the uniform topology. One can see that a von Neumann algebra is also closed under the weak topology.

By definition, given a net $\{A_\alpha\}_{\alpha\in\mathcal I}\subset\mathcal B(\mathcal H)$ of bounded operators, we say that it converges weakly to $A\in\mathcal B(\mathcal H)$ if
\begin{equation}
	A_\alpha\xrightarrow{w} A\iff \forall \varphi,\psi\in\mathcal H:\ \braket{\varphi}{ A_\alpha\psi}\rightarrow \braket{\varphi}{ A\psi}.
\end{equation}
Von Neumann algebras are closed with respect to the topology induced by this notion of convergence.

One of the fundamental results regarding von Neumann algebras is the bicommutant theorem~\cite{bratteli2012operator,Neumann1930}:
\begin{thm}[Bicommutant theorem]\label{th:bicommutantheorem}
	A nondegenerate algebra $\mathfrak M\subset\mathcal B(\mathcal H)$ is a von Neumann algebra if and only if it is closed in the weak topology.
\end{thm}

Other topologies on $\mathcal B(\mathcal H)$ could have been considered ($\sigma$-weak, strong, $\sigma$-strong, strong$^*$ and $\sigma$- strong$^*$), and for all of them Theorem~\ref{th:bicommutantheorem} works, see~\cite{bratteli2012operator}.

Given a $*$-algebra of operators, it is always possible to obtain a von Neumann algebra by considering its bicommutant, and it would be exactly the weak closure of the algebra itself. In a seminal paper~\cite{onringoperator}, Murray and von Neumann  introduced a classification of von Neumann algebras. The main motivation of this work was characterizing all  possible algebras satisfying the condition
\begin{equation}\label{eq:factordefinition}
	\mathfrak M\cap \mathfrak M'=\mathbb C\mathbb I\,.
\end{equation}
In a quantum mechanics framework, $\mathfrak M$ and $\mathfrak M'$ can be interpreted as algebras of independent systems.  An algebra satisfying condition~\eqref{eq:factordefinition} is called a factor. For example, given a tensor product of Hilbert spaces
\begin{equation}
	\mathcal H=\mathcal H_1\otimes \mathcal H_2,
\end{equation}
the algebras $\mathcal B(\mathcal H_1)\otimes \mathds I_2$ and $ \mathds I_1 \otimes \mathcal B(\mathcal H_2)$ are examples of factors.
Von Neumann's original problem consisted in finding factors which do not appear in this simple form.

In~\cite{onringoperator}, factors were sorted into three different types via a dimension function $\mathfrak D$. This characterization can be extended and generalized to all von Neumann algebras. We here summarize this characterization, see~\cite{TakesakiI} for  details. Before doing this, we need to define an equivalence relation between projections.
\begin{definition}
	Let $\mathfrak M\subset \mathcal B(\mathcal H)$ be a von Neumann algebra, and let $p$, $q$ be projections in $\mathfrak M$. We say that $p$ is similar to $q$, and we write $p\sim q$, if there exists a linear partial isometry $U\in\mathfrak M$ such that $p=U^*U$ and $q=UU^*$.
\end{definition}
Note that, under the previous assumption, $U$ is an isometry between the support of $p$ and the support of $q$.
\begin{definition}
	Let $\mathfrak M$ be a von Neumann algebra, and call its center $Z=\mathfrak M\cap \mathfrak M'$. Then
	\begin{enumerate}
		\item\label{item:type_i} $\mathfrak M$ is of \textbf{Type} $\boldsymbol{I}$ if every non-zero central projection $z\in Z$ has an abelian sub-projection $p\leqslant z$ in $\mathfrak M$. By definition, this means that the sub-algebra $p\mathfrak M p$ is an abelian $W^*$-algebra on $p\mathcal H$.
		\item\label{item:type_ii} $\mathfrak M$ is of \textbf{Type} $\boldsymbol{II}$ if it does not have any abelian projection. Moreover, every non-zero central  projection $z\in Z$ 
		majorizes a nonzero-finite projection $p\leqslant z$ in $\mathfrak M$. Here, a projection $p$ is finite if there are no proper sub-projections $q<p$ similar to $p$.
		\item\label{item:type_iii} $\mathfrak M$ is of \textbf{Type} $\boldsymbol{III}$ if it does not have any finite projection.
	\end{enumerate}
\end{definition}
Note that the notion of finiteness in~\ref{item:type_ii}. is standard in set theory.

In general, every von Neumann algebra can be uniquely decomposed  as the direct sum of type $I$, $II$ and $III$ algebras~\cite{TakesakiI}.
Type $I$ algebras are the ones used in standard quantum mechanics. If they are factors, they turn out to be isomorphic to complete algebras of operators over a Hilbert space $\mathcal B(\mathcal H)$. In the general case, they are just the direct sum (or integral) of such complete algebras.

Type $II$ algebras can be subdivided into two kinds. In particular, they are type $II_{1}$ if the identity if finite, and $II_\infty$ if there are no finite central projection $z\in Z$. Murray and von Neumann proved that all algebras of type $II_1$ also possess a unique finite trace. In this sense, they can be interpreted as the continuum extensions of finite matrix algebras. Moreover, they proved that there exists a unique hyperfinite type $II_1$ factor, up to isomorphism.

Type $III$ algebras, which were introduced in an abstract and mathematical framework, turned out to be the relevant ones in quantum fields theory~\cite{Haag1996,Fredenhagen1985}. In his seminal work~\cite{Powers1967}, Powers for the first time constructed a continuum family of nonequivalent type $III$ factors, proving that there is an uncountable family of algebras of this type. In particular, his work was then generalized in~\cite{arakiclassification}, considering more general families of matrix algebras than $2\times2$ (see below). Later on,  Krieger~\cite{krieger,Krieger1976} and then Connes \textit{et al.}\cite{connesfeldmanweiss,connes1995noncommutative} characterized a broader family of algebras, which is the family of amenable algebras.

Let us also anticipate that the same construction from Powers, which depends on a parameter $\lambda$, can also be used in order to get the hyperfinite type $II_1$ factor. In the final part of this section, we will give a quick summary of this construction. It will be relevant, as we are going to show how it is equivalent to the algebra of observables of a properly defined groupoid.

Consider the $C^*$-algebra $\mathfrak{A}$ obtained by induction from the tensor products of $2\times 2$ matrix algebras. Explicitly, take the increasing sequence
\begin{equation}
	\label{eq:tensor_product_matrix_algebras}
	\mathcal A_n=M_{2^n}=\bigotimes_{k=1}^{n}M_2
\end{equation}
with the embedding $A\in \mathcal A_n \mapsto  A\otimes \mathbb I_2\in\mathcal A_{n+1}$. Call $\mathfrak{A}_0=\bigcup_n \mathcal A_{n}$ the union of the finite dimensional algebras $\mathcal A_{n}$, and  $\mathfrak{A}$ the $C^*$-algebra generated by this union:
\begin{equation}\label{eq:powercstaralgebra}
	\mathfrak A=\overline{\bigcup_{n\in\mathbb N}  \mathcal A_{n}}=\overline{\mathfrak A_0}.
\end{equation}
In particular, this $C^*$-algebra is defined as the closure  of $\mathfrak A_0$ in the norm inherited from the uniform norm of $ \mathcal A_n$. Consider a sequence of density matrices $(\rho_n)_{n\in\mathbb N}$ in $M_{2}$, and define the corresponding tensor product state on $\mathfrak A$:
\begin{equation}
	\label{eq:state_tensor_product}
	\phi=\bigotimes_{n\in\mathbb N} \rho_n.
\end{equation} 
To be more precise, we can define $\phi$ on each $\mathcal A_n$ as a finite tensor product
\begin{equation}
	\phi(A)=\Tr((\otimes_{k=1}^n\rho_k) A),\qquad A\in\mathcal A_n.
\end{equation}

This will define $\phi$ on $\mathfrak A_0$. Being $\phi$ norm continuous, it will extend uniquely to a state on $\mathfrak A$. In particular, Powers considered the special case where all $\rho_n$ are equal to a certain $\rho_\lambda$, which is given by

\begin{equation}
	\label{eq:powerstateproduct}
	\rho_\lambda =
	\begin{pmatrix}
		\lambda & 0\\
		0& 1-\lambda
	\end{pmatrix}
\end{equation}
and $0\leqslant\lambda\leqslant 1/2$.
Then, he considered the \textit{GNS}-representation corresponding to this state, which we will indicate as $(\mathcal H_\lambda,\pi_\lambda,\Psi_\lambda)$, and the von Neumann algebra $\pi_\lambda(\mathfrak A)''$ generated by $\pi_\lambda(\mathfrak A)$.

It turns out that for every value of $\lambda$ the corresponding von Neumann algebra $\mathfrak{M}_{\lambda}$ is a factor. In particular, type $III$ factors are obtained for $0<\lambda<1/2$, whereas for $\lambda=1/2$, one gets the hyperfinite factor $II_1$. Eventually, he proved that different values of $\lambda$ generate non-isomorphic factors, showing that there is a continuous family of non-equivalent type $III$ factors.

Interestingly, it is possible to obtain this cyclic representation by introducing infinite tensor product, which is the way in which von Neumann originally worked~\cite{von1939infinite}. Unfortunately, he could not frame it as a GNS-representation, the latter being introduced afterwords.

Finally, observe that the approach from Araki-Woods~\cite{arakiclassification} generalizes the previous discussion by assuming the algebra $\mathfrak A$ to be defined from matrix algebras of different dimension, and assuming different $\rho_n$ in equation~\eqref{eq:state_tensor_product}.

\subsection{Left-Hilbert Algebras}\label{ch:lefthilbertalgebras}
Left-Hilbert algebras have been introduced by Tomita~\cite{Tom67a}, as the natural framework for modular theory, and then they were revisited by Takesaki~\cite{TakesakiII,Takesaki1970}. They have an important role in thermodynamics, as they clarify the connection between dynamics and equilibrium states~\cite{bratteli2012operator,Kubo,MartinSchwinger,haagtd}. In particular, they can be employed in the characterization of the asymptotic dynamics of an open quantum system, see~\cite{Longo2020} for the continuum time case, and~\cite{AFK_asympt} for the discrete one. Clearly, they also play an important role in the classification of factors~\cite{connes1995noncommutative,connes_frencharticle}. In the Schwinger's picture of quantum mechanics, they take a central role since the algebra of observables is constructed as the von Neumann algebra generated by a certain left-Hilbert algebra~\cite{lgcqt,hahnhaarmeasure,hahnregularrep}.

\begin{definition}[Left Hilbert algebras]
	A left-Hilbert algebra is a complex algebra $\mathfrak C$ together with an inner product $\braket{\cdot},{\cdot}$ (under which it is not complete) and an involution $x\rightarrow x^\#$ with the following properties:
	\begin{enumerate}
		\item[(i)] For all $\xi \in \mathfrak C$, the map $\pi_\ell(\xi):\mathfrak C \rightarrow\mathfrak C$, $\eta\mapsto\xi\eta$ is bounded;
		\item[(ii)] For all $\xi,\eta$ and $\zeta$ in $\mathfrak C$, one has $\braket{\zeta}{\xi\eta}=\braket{\xi^{\#}\zeta}{\eta}$, or equivalently $\pi_\ell(\xi)^\dagger=\pi_\ell(\xi^\#)$;
		\item[(iii)] The involution $\#$ is closable in $\mathcal H$, $\mathcal H$ being the closure of $\mathfrak C$ under the inner product topology;
		\item[(iv)] $\mathfrak C^2 = \left\lbrace \xi\eta, \xi,\eta \in \mathfrak{C}  \right\rbrace$ is dense in $\mathcal H$.
	\end{enumerate}
\end{definition}
A simple example is the cyclic representation $(\mathcal H_\omega,\pi_{\omega},\Psi_\omega)$ of a $C^*$-algebra $\mathfrak A$ associated with a state $\omega$. Suppose that the state $\Psi_\omega$ is separating for $\pi_\omega(\mathfrak A)$, namely
\begin{equation}
	A\Psi_\omega=B\Psi_\omega\implies A=B,
\end{equation}
with $A,B$ in $\pi_\omega(\mathfrak A)$. Then, a left-Hilbert algebra is obtained by considering
\begin{equation}
	\mathfrak C=\{\pi_\omega(A)\Psi_\omega|A\in\mathfrak A\}\subset \mathcal H_\omega
\end{equation}
with a product and an involution respectively given by
\begin{align}
	(\pi_\omega(A)\Psi_\omega) (\pi_\omega(B)\Psi_\omega)&=\pi_\omega(AB)\Psi_\omega,\\
	(\pi_\omega(A)\Psi_\omega)^\#&=\pi_\omega(A)^*\Psi_\omega=\pi_\omega(A^*)\Psi_\omega.
\end{align}

Similarly to a left-Hilbert algebra, it is possible to define a right-Hilbert algebra. In this case, $\pi_\ell$ in $(i)$ is replaced by the operator $\pi_r(\eta)(\xi)=\xi\eta$, which must be bounded, and the involution, indicated as $\eta\mapsto\eta^{\flat}$, must satisfy $\braket{\zeta}{\xi\eta}=\braket*{\zeta\eta^{\flat}}{\xi}$ instead of $(ii)$.

It is possible to obtain a right-Hilbert algebra from a left-Hilbert algebra. In particular, one has to define a right bounded product and an involution $\flat$. The involution $\flat$ is obtained as the adjoint of the involution $\#$. To be more precise, call
\begin{align}
	S:\mathcal D^{\#}\subset \mathcal H \rightarrow \mathcal D^{\#} 
\end{align}
the closure of $\#$, and $F$ its adjoint
\begin{equation}
	F:\mathcal D^{\flat}\subset \mathcal H \rightarrow \mathcal D^{\flat} ,
\end{equation}
satisfying, for all $\xi \in \mathcal D^{\#}$ and $\eta\in\mathcal D^{\flat}$
\begin{equation}
	\braket{\eta}{S\xi}=\braket{\xi}{F\eta}.
\end{equation}
Note that $F$ and $S$ are antilinear operators. In particular, one can define the modular operator $\Delta=FS$, and obtain the polar decomposition of $S$:
\begin{equation}\label{eq:TTpolardecomposition}
	S=J\Delta^{1/2},
\end{equation}
with the antiunitary operator $J$ called modular conjugation. As we shall see, the modular operator determines a dynamic on the algebra which can be considered intrinsic to the system.

The right product is obtained by considering all the elements $\eta$ in $\mathcal H$ for which the product $\pi_\ell(\xi)\eta$ is bounded in $\xi$. Call $\mathfrak B'$ the set of all such $\eta$, and set $\pi_r(\eta)\xi=\pi_\ell(\xi)\eta$. Then, one can prove that the set
\begin{equation}
	\mathfrak C'=\mathfrak B'\cap \mathcal D^{\flat}
\end{equation}
is actually a right-Hilbert algebra. Here, a right product and an involution are clearly defined. The relation between left-Hilbert algebra and right-Hilbert algebra is very similar to the relation between a von Neumann algebra and its commutant. In particular, if we call $\mathcal R_\ell(\mathfrak C)$ the von Neumann algebra generated by $\pi_\ell(\mathfrak C)$, and $\mathcal R_r(\mathfrak C')$ the one generated by $\pi_r(\mathfrak C')$, one has
\begin{equation}
	\mathcal R_r(\mathfrak C')=\mathcal R_r(\mathfrak C)'.
\end{equation}
Repeating a similar argument for $\mathfrak C'$, one can obtain another left-Hilbert algebra $\mathfrak C''\supset \mathfrak C$, and so on:
\begin{align}
	\mathfrak C&\subset\mathfrak C''=\mathfrak C^{iv}\dots,\\
	\mathfrak C&'=\mathfrak C'''=\mathfrak C^{v}\dots.
\end{align}
A left-Hilbert algebra is said to be full whenever $\mathfrak C=\mathfrak C''$.

The main result regarding left-Hilbert algebras and modular theory is connected with the automorphism of the algebra naturally arising from it.

\begin{thm}[Tomita-Takesaki theorem]
	\label{th:TTtheorem}
	Let $\mathfrak C$ be a left-Hilbert algebra, with  modular operator $\Delta$ and  modular conjugation $J$. Call $\mathcal R_\ell (\mathfrak C)$ the von Neumann algebra generated by $\mathfrak C$. The following relations hold:
	\begin{align}
		J\mathcal R_\ell (\mathfrak C)J=\mathcal R_\ell (\mathfrak C)',
		\\
		J\mathcal R_\ell (\mathfrak C)'J=\mathcal R_\ell (\mathfrak C),
		\\
		\Delta^{\mathrm i t}\mathcal R_\ell (\mathfrak C)\Delta^{-\mathrm i t}=\mathcal R_\ell (\mathfrak C),
		\label{eq:modulardynamics}
		\\
		\Delta^{\mathrm i t}\mathcal R_\ell (\mathfrak C)'\Delta^{-\mathrm i t}=\mathcal R_\ell (\mathfrak C)'.
	\end{align}
\end{thm}
The first two relations connect the von Neumann algebra $\mathcal R_\ell (\mathfrak C)$ with its commutant via the operator $J$. The second ones represent a unitary dynamics (with respect to the inner product) for the algebra, generated by the Hamiltonian $H=\log \Delta$.

\section{Groupoids and groupoid algebras}
\label{sec:groupoid_and_groupoid_algebras}
In this section, we are going to summarize the main definitions of groupoid, and the algebra of observables associated with it. For details, see~\cite{groupoidI,groupoidII,ibort2019introduction,hahnhaarmeasure,hahnregularrep,groupoidIII,groupoidIV,landsman2012mathematical}.

A groupoid generalizes the concept of group as not all of its elements can be composed. With the terminology from Category theory, a groupoid is a set $\mathcal G$, which we will call the set of transitions, together with a set of objects $\Omega$, the so-called base. Here, two maps $t$ (target) and $s$ (source) from $\mathcal G$ onto $\Omega$ exist, which characterize the product (or composition) $\alpha\circ\beta$ between two elements $\alpha$, $\beta$ of the groupoid $\mathcal G$ in the following way.

Let $\alpha,\ \beta$ and $\gamma$ be in $\mathcal G$, and let $x\in \Omega$.
\begin{enumerate}
	\item[(i)] $\alpha\circ\beta$ is defined if and only if $s(\alpha)=t(\beta)$, in which case $s(\alpha\circ\beta)=s(\beta)$ and $ t(\alpha\circ\beta)=t(\alpha)$.
	\item[(ii)] There is an identity $\mathbb I_x\in \mathcal G$ for each $x\in\Omega$, with $s(\mathbb I_x)=t(\mathbb I_x)=x$, and such that $\alpha\circ\mathbb I_x=\alpha$ and $\mathbb I_x \circ \beta=\beta$, for all $\alpha, \beta$ satisfying $x=s(\alpha)=t(\beta)$.
	\item[(iii)] Associativity: $(\alpha\circ\beta)\circ\gamma=\alpha\circ(\beta\circ\gamma)$ whenever one of the two sides is well defined.
	\item[(iv)] Any $\alpha\in\mathcal G$ has a two-side inverse $\alpha^{-1}$, namely $\alpha^{-1}\circ \alpha=\mathbb I_{s(\alpha)}$ and $\alpha\circ\alpha^{-1} =\mathbb I_{t(\alpha)}$.
\end{enumerate}
Using the category theory formalism, we write
\begin{equation}
	\alpha:x\rightarrow y,
\end{equation}
to indicate that $\alpha\in\mathcal G$ has $s(\alpha)=x$ and $t(\alpha)=y$. Inspired by Schwinger's approach to quantum mechanics, we interpret the base $\Omega$ as the set of outcomes of a measurement. In this way, an $\alpha$ in $\mathcal G$ can be interpreted as a transition from the outcome $x$ to the outcome $y$.

Analogously to the way one introduces the group-algebra of a group, we will now describe how to define a groupoid-algebra. This will be interpreted as the algebra of physical observables, connecting the groupoid picture with the algebraic picture of quantum mechanics. To better understand these notions, we start with the finite case.

\begin{example}[Example: finite groupoids]
	\label{ex:finite_groupoids}
	Consider a finite groupoid $\mathcal G=\{\gamma_k|k=1,\dots,N\}$, with base $\Omega=\{x_a|a=1,\dots,n\}$, $n\leqslant N$.
	We can define the groupoid algebra $\mathbb C[\mathcal G]$, in the following way
	\begin{equation}\label{eq:finitedimensionalalgebra}
		\mathbb C[\mathcal G]=\mathrm{span}(\mathcal G)=\left\{\sum_{k=1}^N{\textbf A=A_k\gamma_k}|A_k\in \mathbb C\right\}
	\end{equation}
	and the product between two elements is given by
	\begin{equation}\label{eq:productfinitedimensional}
		\textbf{A} \star\textbf{B}=\sum_{k,j=1}^N A_k B_l \gamma_k \circ \gamma_j.
	\end{equation}
	The involution of an element in the algebra $\textbf A=\sum_{k=1}^N {A_k} \gamma_k$ is defined as
	\begin{equation}
		\textbf A^\dagger=\sum_{k=1}^N \overline{A_k} \gamma_k^{-1}
	\end{equation}
	
	Note that this algebra is noncommutative as the composition law of the groupoid is not. The groupoid algebra corresponds to the algebra of observables in the algebraic description of quantum mechanics, and its self-adjoint elements are the observables. By considering the product~\eqref{eq:productfinitedimensional} as an operator $\textbf A\star$ acting on $\textbf B\in\mathbb{C}[\mathcal G]$, we get a natural representation of the algebra, which is called the regular representation. When we switch to infinite groupoids, this will became a left-Hilbert algebra.
	
	Another representation is obtained by considering the Hilbert space
	\begin{equation}
		\mathcal H_{\Omega}=\left\{\varphi=\sum_{a=1}^n\varphi_a \ket{x_a}|\varphi_a\in \mathbb C\right\}.
	\end{equation}
	Essentially, the elements (outcomes) of $\Omega$ define a basis for the Hilbert space $\mathcal H_\Omega$. Then, we can define a representation $\pi$ of $\mathbb C[G]$ on $\mathcal H_\Omega$ by assigning to each element in the groupoid $\gamma:x\mapsto y$ the operator
	\begin{align}\label{eq:regularrepresentationgeneralformula}
		\gamma\in G \mapsto\ketbra{y}{x}\in\mathcal B(\mathcal H_\Omega).
	\end{align}
	This is called the fundamental representation of the groupoid algebra. One obtains the standard interpretation of the elements of groupoids as rank-one operators between vectors associated with the outcomes of a measurement. Every transition in an isotropy group (the set denoted $\mathcal{G}_x^x$ of transitions $\gamma\in \mathcal{G}$ such that $s(\gamma)=t(\gamma)=x$ is a group and we call it the isotropy group at $x$) is represented as a rank-one projector $\ketbra*{x}$.
\end{example}

\begin{example}[Example: groupoid of pairs] \label{example:standardcouplegroupoid}
	Consider a finite groupoid $\mathcal G = \Omega\times \Omega$ which is just the direct product of the base set $\Omega$ with itself:
	\begin{equation}
		\mathcal G=\Omega\times\Omega.
	\end{equation}
	We set, for $(x,y)\in\mathcal G$, the source and target functions to be
	\begin{equation}
		t(x,y)=x,\quad s(x,y)=y.
	\end{equation}
	The composition of $(x,y)$ and $(w,z)$ is possible iff $y=w$, in which case $(x,y)\circ(y,z)=(x,z)$. Then, the groupoid algebra becomes
	\begin{equation}
		\mathbb{C}[\Omega\times\Omega]=\left\{\textbf c=\sum_{xy} c_{xy}(x,y)|x,y\in \Omega,c_{xy}\in\mathbb C\right\}.
	\end{equation}
	Using the fundamental representation, this is nothing but the complete algebra of matrices over $\mathbb{C}[\Omega]$. Identifying $(x,y)\leftrightarrow \ketbra{x}{y}$, the $\star$ product of two elemets $\textbf{c}=\sum_{xy}c_{xy}\ketbra{x}{y}$ and $\textbf{d}=\sum_{xy}d_{xy}\ketbra{x}{y}$ is
	\begin{equation}
		\textbf{c}\star\textbf{d}=\sum_{xyz}c_{xy}d_{yz}\ketbra{x}{z},
	\end{equation}
	which is just the matrix multiplication between the matrices $[c_{xy}]$ and $[d_{xy}]$. We get standard quantum mechanics, with the algebra of observables given by all the operators on ${\mathbb C}[\Omega]$:
	\begin{equation}\label{eq:standardalgebra}
		\nu[\Omega\times\Omega]=\mathcal B(\mathcal H_\Omega)=M_n
	\end{equation}
\end{example}

A couple of comments are in order. First of all, for generic groupoids the fundamental representation is not faithful, since the isotropy groups are trivially represented as the same element, see the last line of Example~\ref{ex:finite_groupoids}. 
Therefore, the fundamental representation of the groupoid of pairs of a set $\Omega$ corresponds to the standard case of a Hilbert space in which one has selected a complete family of compatible observables and no gauge degrees of freedom are considered. Generic groupoids having non-trivial isotropy groups allow some gauge degrees of freedom to enter the description. These redundant degrees of freedom, however, disappear in the fundamental representation. We will see that a faithful representation can be obtained in a different way, via the regular representation. 

\subsection{Groupoid algebras}\label{ch:groupoidalgebras}
In this subsection we present the construction of the groupoid-algebra and its regular representation for measure groupoids, i.e., groupoids together with their Borel algebras of measurable subsets and a special equivalence class of measures. The reason for the choice of a measure is the need of replacing the sum in~\eqref{eq:finitedimensionalalgebra} with an integral, whenever non-discrete groupoids are involved. Similarly to the group case~\cite[p.~120]{loomis2013introduction}, this measure will need to be a generalized left-invariant Haar measure. Using such measure, it is possible to obtain a continuous regular representation of the theory. Here, we will just summarize the main results, without dwelling too much on the formalism. For details, see~\cite{lgcqt,hahnhaarmeasure,hahnregularrep}.

Consider the pair $(\mathcal G, \left[ \mu \right])$ consisting of a measurable analytic groupoid $\mathcal{G}$ and an equivalence class of measures $\left[ \mu \right]$, and suppose that a probability $\lambda$ belongs to this class of measures. Since $\lambda$ is a probability measure, the induced measure
\begin{equation}\label{eq:pushbackdellaprobabilita}
	\tilde \lambda=t_{*}\lambda=\lambda\circ t^{-1}\,,
\end{equation}
is well defined. Let $\mu$ be a measure on $\mathcal G$, which is equivalent to $\lambda$. Using the disintegration theorem with respect to the target map $t\,\colon \,\mathcal{G}\,\rightarrow\,\Omega$, it can be decomposed with respect to $\tilde\lambda$ as
\begin{equation}\label{eq:decompositionmeasureorthogonal}
	\mu=\int_{\Omega} \mu^x d\tilde\lambda(x),
\end{equation}
where $\mu^x(\mathcal G \setminus t^{-1}(\{x\}))=0$ for all $x\in \Omega$~\cite{hahnhaarmeasure,Halmos1941,effros}. In particular, this decomposition is unique and the family of measures $\{\mu^x\}_{x\in\Omega}$ is uniquely determined up to a zero-measure set on $\Omega$, which implies that this decomposition is actually associated with the whole equivalence class of measures.

Given $\mu$, the family $\{\mu^x\}_{x\in\Omega}$ is said to be a system of left-invariant Haar measures if it is invariant under the left composition
\begin{equation}\label{eq:lefthaarinvariance}
	(L_\alpha)_* \mu^{s(\alpha)}=\mu^{s(\alpha)}\circ L_\alpha^{-1}=\mu^{t(\alpha)},
\end{equation}
where
\begin{align}\label{eq:leftaction}
	L_\alpha:\mathcal G^{s(\alpha)}&\rightarrow \mathcal G^{t(\alpha)}\nonumber\\
	\beta&\mapsto \alpha\circ\beta.
\end{align}
Here, $G^x=\{\alpha\in G|t(\alpha)=x\}$.
In a nutshell, the family of measures $\{\mu^x\}_{x\in\Omega}$ is not modified by composition on the left. In~\cite{hahnhaarmeasure} it is proved that any measure groupoid $(\mathcal G,\lambda)$, with $\lambda$ a probability invariant with respect to inversion (see below), admits a family of left-invariant Haar measures, obtained by decomposing a measure $\mu$ equivalent to $\lambda$. 
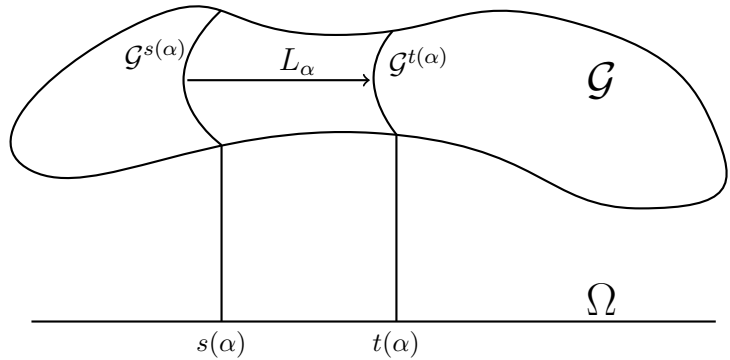
\begin{figure}[hbtp]\center
	\begin{tikzpicture}
		
		\draw [thick] plot [smooth cycle, tension = 1] coordinates {(-3,0) (-1.5,2) (1,1.8) (4,2) (6,0.5) (5,-0.5) (1.5,0.5)};
		
		\draw[thick](-3,-2)--(6,-2);
		
		\draw[thick] plot [smooth, tension=1] coordinates {(-.5,2.129) (-1,1.2) (-.5,0.34)};

		\draw[thick] plot [smooth, tension=1] coordinates {(1.75,1.85) (1.5,1.2) (1.8,0.48)};
		
		\draw[thick](-.5,0.34)--(-.5,-2);
		
		\draw[thick](1.8,0.48)--(1.8,-2);
		
		\draw[thick,->] (-0.95,1.2) -- (1.45,1.2);
		
		\node at (0.5,1.45) {{$L_{\alpha}$}};
		
		\node at (2.1,1.45) {\small{$\mathcal G^{t(\alpha)}$}};
		
		\node at (-1.36,1.55) {\small{$\mathcal G^{s(\alpha)}$}};
		
		\node at (-.5,-2.3) {\footnotesize{$s(\alpha)$}};
		
		\node at (1.8,-2.3) {\footnotesize{$t(\alpha)$}};
		
		\node at (4.5,1.2) {\Large{$\mathcal G$}};
		
		\node at (4.5,-1.7) {\Large{$\Omega$}};
		
	\end{tikzpicture}
	\caption{The left action $L_\alpha$ of an element of the groupoid $\alpha$ will map the subset $\mathcal G^{s(\alpha)}$ of elements which can be composed with $\alpha$, to $\mathcal G^{t(\alpha)}$, whose elements can be composed with $\alpha^{-1}$. An Haar measure $\mu^{s(\alpha)}$ will transform accordingly.}
\end{figure}

Using a family of Haar measures, it is possible to construct an involutive algebra, which allows to represent the groupoid as a family of operators, and extend the concept of groupoid algebra~\eqref{eq:finitedimensionalalgebra} to the continuous case. This algebra will play the role of the algebra of observables in the Schwinger picture.

Let $(\mathcal G,\lambda)$ be a measure groupoid, and consider a measure $\mu$ absolutely continuous with respect to $\lambda$, and $\sigma$-finite. Write this measure as
\begin{equation}\label{eq:inversefunction}
	\mu=\int_{\Omega} \mu^x d\tilde\lambda(x),
\end{equation}
where $\tilde\lambda$ is defined as~\eqref{eq:pushbackdellaprobabilita}, and suppose that the family $\{\mu^x\}_x$ is a system of Haar measures, namely, it satisfies equation~\eqref{eq:lefthaarinvariance}. Again, such a system exists under proper hypotheses on $\lambda$.

Moreover, suppose that the measure $\mu$ is similar to its inverse. This means that, calling the inverse
\begin{align}
	\tau:\mathcal G&\rightarrow \mathcal G\nonumber\\
	\alpha & \mapsto \alpha^{-1},
\end{align}
the push-back of $\mu$ with respect to $\tau$ is similar to $\mu$:
\begin{equation}\label{eq:radonnikodinderivativeinverse}
	\tau_*\mu=\mu\circ\tau^{-1}=\Delta^{-1} \mu.
\end{equation}
Here, $\Delta^{-1}$ is a function from $\mathcal G$ to $\mathbb R_{+}$, and represents the Radon-Nikodym derivative of the inverse measure $\tau_*\mu$ with respect to the measure $\mu$. We will call $\Delta$ modular function, and we shall see that it is connected to the Tomita-Takesaki modular operator discussed in Section~\ref{ch:lefthilbertalgebras}. It can be proved that 
\begin{align}
	\Delta(\alpha^{-1})&=\Delta(\alpha)^{-1},\nonumber \\
	\Delta(\alpha\circ\beta)&=\Delta(\alpha) \Delta(\beta)\,,
\end{align}
which means that the modular function is a groupoid homomorphism from the groupoid $\mathcal{G}$ to the multiplicative group $\mathbb{R}_+$.

It is possible to construct an algebra starting from this measure. Let $F$ be a measurable function
\begin{equation}
	F:\mathcal G\rightarrow \mathbb C,
\end{equation}
and define its Hahn  norm as
\begin{equation}\label{eq:hanhnorm_generalformula}
	\norm{F}_\mathrm{L^{1}_H}=\max\left\{\norm{\int_{\mathcal G}\abs{F(\alpha)}d\mu^{x}(\alpha)}_{\mathrm{L}^{\infty}}, \norm{\int_{\mathcal G}\Delta^{-1}(\alpha)\abs{F(\alpha^{-1})}d\mu^{x}(\alpha)}_{\mathrm{L}^{\infty}}\right\}.
\end{equation}
Here, ${\mathrm L^\infty}$ is the essential supremum over the space $\Omega$, with respect to the measure $\tilde\lambda$ and the parameter $x$ in the integrals. Call $\mathrm{L^{1}_H}(\mathcal G)$ the set of functions with finite Hahn norm. Then, the vector space
\begin{equation}\label{eq:convolutionalgebrageneraldiscussion}
	\mathfrak C =\mathrm{L^{1}_H}(\mathcal G)\cap \mathrm{L^{2}}(\mathcal G)
\end{equation}
is a left-Hilbert algebra~\cite{lgcqt,Tom67a}, with involution and product given respectively by
\begin{align}\label{adjoint_generalformula}
	F^{\dagger}(\alpha)&=\Delta^{-1}(\alpha)\overline{F(\alpha^{-1})}=\Delta(\alpha^{-1})\overline{F(\alpha^{-1})},\\
	(F\star G)(\alpha)&=\int_{\mathcal G} F(\beta) G(\beta^{-1}\circ\alpha) d\mu^{t(\alpha)}(\beta).
	\label{eq:convolution_generalformula}
\end{align}

The von Neumann algebra generated by $\mathfrak C$ will be indicated by $\nu[\mathcal G]=\mathfrak C''$, and it will represent the algebra of physical observables. It is clear that in the finite case, using the counting measure, the previous equations will yield the algebra~\eqref{eq:finitedimensionalalgebra}.

In general, the algebra $\nu[\mathcal G]$ is non commutative, and in this sense the theory we are describing is non-classical. This representation is not irreducible but provides an algebra already in a standard form. According to the general theory of left-Hilbert algebras, the involution operator $\dagger$ can be written as the product $J\Delta^{1/2}$ where $J$ is an antiunitary operator and $\Delta$ is the modular operator. In this case the operator $\Delta\,\colon\,\mathfrak{C}\,\rightarrow\,\mathfrak{C}$ is the multiplication operator by the modular function, whereas $J\,\colon\,\mathrm{L^{2}}(\mathcal G)\,\rightarrow\,\mathrm{L^{2}}(\mathcal G)$ acts as follows:
\begin{equation}
	Jf  = \delta^{-\frac{1}{2}}\overline{f\circ \tau}\,.
\end{equation}
This representation of the algebra is not irreducible, and the commutant, obtained by coniugation with the operator $J$, corresponds to the algebra generated by the right action of the groupoid on itself.

A commutative subalgebra corresponds to the measure of outcomes in $\Omega$, at least in the finite case. We infer this algebra from the example of standard groupoid of pairs, Example~\ref{example:standardcouplegroupoid}. Here, the physical algebra corresponding to $\Omega\times\Omega$ can be represented as a complete algebra of matrices, as in equation~\eqref{eq:standardalgebra}. Using standard quantum mechanics, the classical sub-algebra corresponding to the computational base ${\ket{x}}_{x\in\Omega}$ is the algebra of diagonal matrices:
\begin{equation}
	\nu_\mathrm{cl}(\Omega\times\Omega)=\mathrm{diag}({\lambda_x}:x\in\Omega).
\end{equation}
The action of this subalgebra in the regular representation is 
\begin{equation}
	\sum_x \lambda_x \ketbra{x}{x}\star\sum_{yz}c_{yz} \ketbra{y}{z}=\sum_{xz}\lambda_x c_{xz}\ketbra{x}{z}.
\end{equation}

\section{The infinite qubit-chain groupoid}
\label{ch:infinite_qubit_chain_groupoid}
In this section, we are going to apply the construction of the previous section to a particular example of groupoid. We will consider an infinite countable number of copies of the qubit groupoid~\cite{groupoidI}. It can be interpreted as a chain of spins. We are going to show that the physical algebra corresponding to this example can be of either type $II_1$ or $III$, depending on the chosen measure on the groupoid. This shows how nontrivial algebras may naturally arise in the groupoid picture. 

Let us start with a bit of notation. Let $i\in\mathbb N$, and call
\begin{equation}
	\Omega_i=\{0,1\},
\end{equation}
the base of the $i$-th qubit, whose elements are the possible outcomes of a measurament on it. The base of the chain will be
\begin{equation}\label{eq:baseinfinitecouplegroupoid}
	\Omega_\infty=\prod_{i\in\mathbb N} \Omega_i=\{(x_1,x_2,\dots)| x_i\in\Omega_i\}.
\end{equation}
A sequence $(x_1,x_2,\dots)$ in $\Omega_\infty$ will represent a measurement on the whole chain, getting information on each site.

As discussed previously, noncommutative aspects enters the theory via transitions. We will introduce transitions that only allow a finite number of spins (or qubits) to flip. They can be introduced on $\Omega_\infty$ as elements of the following countable set
\begin{equation}\label{eq:transitioninfinitecouplegroupoid}
	\Gamma=\{(x_1,x_2,\dots)\in \Omega_\infty|\exists N\ \textrm{s.t.} \ x_i=0\  \forall i>N\}.
\end{equation}
Given an element $x^o\in \Gamma$, if $x_i=1$ is a member of the sequence equal to $1$, there will be a flip of the $i$-th qubit. To be more precise, we will consider the groupoid
\begin{equation}\label{eq:infinitequbitchaingroupoid}
	\mathcal G =  \Omega_\infty\times \Gamma,
\end{equation}
with target and source functions given by
\begin{align}\label{eq:sourceinfinitecouplegroupoid}
	s(x,x^o)&= x\oplus x^o, \\
	t(x,x^o)&= x,\label{eq:targetinfinitecouplegroupoid}
\end{align}
for $(x,x^o)\in  \Omega_\infty\times\Gamma$. Here, the $\oplus$ symbol stands for sum modulo $2$, and the sum of two sequences is intended as the term by term sum. Let us note, here, that the set $\Gamma$ endowed with the operation $\oplus$ forms a countable Abelian group and the inverse operation coincides with $\oplus$ itself. From now on, we will indicate with $x,y,\dots$ the elements in $\Omega_\infty$ and $x^o,y^o,\dots$ the elements of $\Gamma$. Two elements of the groupoid $\alpha = (x,x^o)$ and $\beta = (y,y^o)$ can be composed iff $y=t(\beta)=s(\alpha)=x\oplus x^o$, in which case 
\begin{equation}\label{eq:compositioninfinitecouplegroupoid}
	\alpha \circ\beta= (x,x^o)\circ(y,y^o)=(x,x^o\oplus y^o).
\end{equation}

Note that description given in~\eqref{eq:compositioninfinitecouplegroupoid} differs from the one corresponding to the groupoid of pairs in Example~\ref{example:standardcouplegroupoid}, as the second term in~\eqref{eq:infinitequbitchaingroupoid} is the difference between target and source, rather than the source itself. In particular, note that the groupoid is not connected, as a transition between different elements in the base can be defined if and only if they are equal after a finite number of terms. Actually, this groupoid can be seen as an action groupoid, with the group $(\Gamma,\oplus)$ acting on the base $\Omega_\infty$, see~\cite{groupoidI}.

Given an element $x$ in the base, its identity is $\mathbb I_x=(x,0)$, while the inverse of an element $(x,x^o)$ in the groupoid is 
$(x,x^o)^{-1}=(x\oplus x^o,x^o)$.

We are interpreting this model as a system of spins, and we are measuring their spin along the $z$ axis, assuming that a finite number of qubits (or spins) can be flipped due to noise or quantum jumps and every transition can in principle be reversed.

In order to construct the von Neumann algebra of observables as defined in Section~\ref{ch:groupoidalgebras}, we need a measure on the groupoid $\mathcal G$ which is absolutely continuous to its inverse (Eq.~\eqref{eq:radonnikodinderivativeinverse}) and left invariant (Eq.~\eqref{eq:lefthaarinvariance}). We explicitly construct it  as a product of measures, one over the base $\Omega_\infty$ and the other on the set of transitions $\Gamma$.

The construction of the measure on $\Omega_\infty$ will follow the standard approach to define a probability on an infinite product in probability theory~\cite{kolmogorov2013foundations,Shiryaev1996,Rao1971}. We briefly summarize it. First, we need to define the measurable sets. Given a set $A_N\subset \prod_{i=1}^N\Omega_i$, define the cylinder on $\Omega_\infty$ with base $A_N$ as the set
\begin{equation}
	\label{eq:definition_cylinder}
	\mathscr{C}(A_N)=\{x\in\Omega_\infty|(x_1,\dots,x_N)\in A_N\}.
\end{equation}
This cylinder is the set of all the sequences with the first $N$ elements in $A_N$. Running $N\in\mathbb N$ and $A_N\subset \prod_{i=1}^N\Omega_i$, we get a family of cylinders in $\Omega_\infty$. We define the $\sigma$-algebra $\Sigma_\infty$ as the $\sigma$-algebra generated by these cylinders. It is then possible to obtain a product probability over $\Sigma_\infty$. Let a probability be defined on each $\Omega_i$ by
\begin{equation}\label{eq:singlesiteprobability}
	\nu_i=\lambda_i \delta_{\{0\}}+(1-\lambda_i)\delta_{\{1\}},\quad 0\leqslant\lambda_i\leqslant 1.
\end{equation}
Here, $\delta_{\{k\}}(A)$ is the delta measure at $k$, with value $1$ when $k$ is in $A$, $0$ otherwise.
Then, there exists a unique probability measure $\nu_\infty=\prod_i\nu_i$ on $\Sigma_\infty$ which acts on the cylinders as the product of the finite product measure of $\nu_i$, namely
\begin{equation}
	\nu_\infty(\mathscr{C}(A_N))=\left\{\prod_{i=1}^N\nu_{i} \right\}(A_N),
\end{equation}
for all $N\in\mathbb N$ and $A_N\subset \prod_{i=1}^N\Omega_i$. This result is referred to as Kolmogorov's theorem~\cite[p.~27]{kolmogorov2013foundations}.

We now need a measure on $\Gamma$. Note that $\Gamma$, as a subset of $\Omega_\infty$, is a measurable set of $0$ measure with respect to $\nu_\infty$ (unless the pathological case $\lambda_i=1$, definitively in $i$). In order to get a nontrivial theory, we have to introduce a different measure on it. Since $\Gamma$ is a countable commutative group under $\oplus$, we will use the (group) Haar measure on it, which is just the counting measure $\#$:
\begin{equation}
	\#(A)=\sum_{x^o\in A} 1,
\end{equation}
defined for all $A\subset \Gamma$. The $\sigma$-algebra on $\Gamma$ is the power set of $\Gamma$, $\mathcal P(\Gamma)$. The measure on $\mathcal G$ is given by the product measure $\mu=\nu_\infty\times\#$, over the product $\sigma$-algebra $\Sigma_\infty\times\mathcal P(\Gamma).$

As underlined in Section~\ref{ch:groupoidalgebras}, the measure $\mu$ has to satisfy certain requirements in order to obtain the physical algebra mentioned in the previous section~\cite{hahnhaarmeasure,hahnregularrep}. In particular, it must be equivalent to its inverse~\eqref{eq:radonnikodinderivativeinverse}, and its restrictions to the fibers of the target map must form a system of left-invariant Haar measures~\eqref{eq:lefthaarinvariance}.

\paragraph{Inverse.}

We need to check that the measure $\mu$ is equivalent to its inverse. Let us consider the inverse map
\begin{align}
	\tau: \mathcal{G}=\Omega_\infty\times \Gamma&\rightarrow \mathcal G=\Omega_\infty\times \Gamma\\
	\alpha=(x,x^o)&\mapsto \alpha^{-1}=(x\oplus x^o,\,x^o).
\end{align}
The inverse measure is obtained as the push-forward of $\mu$ by $\tau$:
\begin{equation}
	\tau_*\mu(K)=\mu(K^{-1})=\mu(\{(x,x^o)^{-1}|(x,x^o) \in K\}),
\end{equation}
with $(x,x^o)^{-1}=(x\oplus x^o,x^o)$. By setting $K|_{x^0}=\{x|(x,x^o)\in K\}$, we get
\begin{equation}
	\mu(K^{-1})= \sum_{x^o\in\Gamma}\nu_\infty(K^{-1}|_{x^o})= \sum_{x^o\in\Gamma} ((L_{x^o})_{\ast}\nu_\infty)(K|_{x^o}).
\end{equation}

Now, we claim that $((L_{x^o})_{\ast}\nu_\infty)=\Delta^{-1} \nu_\infty$, with the modular function $\Delta$ given by 
\begin{equation}\label{eq:modularfunction}
	\Delta^{-1}(x,x^o)= \prod_{i=1}^\infty \left(\frac{\lambda_i}{1-\lambda_i}\right)^{(2 x_i-1)x_i^o}.
\end{equation}
Indeed, we choose a finite family of elements $y_i\in \Omega_i$, with $i=1,\dots, N$, and prove the previous formula for the cylinder $\mathscr C(\{(y_1,\dots,y_N)\})$, then the uniqueness in Kolmogorov theorem will then ensure that the equality holds on the whole $\sigma$-algebra $\Sigma_\infty$.
Explicitly,
\begin{align}
	((L_{x^o})_{\ast}\nu_\infty)(\mathscr C((y_1,\dots,y_n))) &=\nu_\infty (\mathscr C(y_1\oplus x ^o_1,\dots,y_N\oplus x^o_N))\nonumber\\
	&=\prod_{i=1}^N[\lambda_i\delta_{\{0\}}(y_i\oplus x^o_i)+(1-\lambda_i)\delta_{\{1\}}(y_i\oplus x^o_i)]\nonumber \\
	&= \prod_{i=1}^N [\lambda_i f^i_0(x^o_i)\delta_{\{0\}}(y_i)+(1-\lambda_i)f^i_1(x^o_i)\delta_{\{1\}}(y_i)].
\end{align}
Here, $\lambda_i f_0^i$ must have value $\lambda_i$ for $x_i^o=0$, and $1-\lambda_i$ for $x_i^o=1$, while $(1-\lambda_i)f_1^i$ must have value $1-\lambda_i$ for $x_i^o=0$, and $\lambda_i$ for $x_i^o=1$. But these are exactly the values of $\Delta^{-1}(x,x^o)$. Compare this equation with~\cite[p.~139]{pukanski_some}. Since $\Delta$ is a strictly positive function, we have that the two measures $\mu$ and $\tau_* \mu$ are absolutely continuous to each others.

\paragraph{Left-invariance.}

Following the approach from~\cite{hahnhaarmeasure,hahnregularrep}, we need to decompose the measure $\mu$ as in~\eqref{eq:inversefunction}, in terms of a probability $\lambda$ absolutely continuous to $\mu$. Consider a probability distribution $m$ over $\Gamma$ in the form
\begin{equation}
	m(A) = \sum_{x^o\in A}m(x^o)\quad 0<m(x^o)<1,\ \sum_{x^o \in \Gamma}m(x^o)=1.
\end{equation}
Then, the probability $\lambda=\nu_\infty\times m$ makes $(\mathcal G,\lambda)$ a measure groupoid. Observe that $\lambda$ is absolutely continuous with respect to $\mu$. Write the push-forward of $\lambda$ with respect to the target function $t$, as in equation~\eqref{eq:pushbackdellaprobabilita}:
\begin{equation}
	\tilde{\lambda}(A)=t_*\lambda(A)=\lambda(t^{-1}(A))=\lambda(A\times \Gamma )=\nu_\infty(A),
\end{equation}
for all $A$ measurable subsets of $\Omega_\infty$. Then
\begin{equation}
	\mu=\int_{\Omega_\infty} d\tilde{\lambda}(x)\mu^x=
	\int_{\Omega_\infty}d\tilde{\lambda}(x)\delta_{\{x\}}\times\# ,
\end{equation}
where
\begin{equation}\label{eq:measuredecomp_infinitecouplegroupoid}
	\mu^{x}=\delta_{\{x\}}\times \#,
\end{equation}
satisfies the property that $\mu^{x}(G\setminus t^{-1}(\{x\})= 0$, see equation~\eqref{eq:decompositionmeasureorthogonal}. We should now check that the family $\{\mu^x\}_{x\in\Omega_\infty}$ is left-invariant. For $\alpha=(x,x^o)$, the left action~\eqref{eq:leftaction} becomes
\begin{align}
	L_\alpha:\{x\oplus x^o\}\times\Gamma&\rightarrow \{x\}\times\Gamma\nonumber \\
	\beta=(x\oplus x^o,y^o) &\mapsto\alpha\circ\beta=(x,x^o\oplus y^o).
\end{align}
Here, $\{x\oplus x^o\}\times\Gamma =t^{-1}(\{x\oplus x^o\})$ is the set of all transformations that can be composed with $\alpha$, while $\{x\}\times\Gamma=t^{-1}(\{x\})$ is the set of transformations with same target as $\alpha$.

Equation~\eqref{eq:lefthaarinvariance} now becomes, for a set $\mathcal{K}\subset\mathcal G= \Omega_\infty\times \Gamma$ such that $K_x=\mathcal{K}\cup t^{-1}(x)\neq \emptyset$:
\begin{align}
	\left( L_{\alpha} \right)_*(\mu^{s(\alpha)})(\mathcal{K})&=\mu^{s(\alpha)}\circ L_{\alpha^{-1}}(\mathcal{K})=\mu^{s(\alpha)}\circ L_{\alpha^{-1}}\left(K_x\right)\nonumber \\
	&=\mu^{x+x^o}\circ L_{(x,x^o)}^{-1}( K_{x})=\#(K_{x})=\mu^x(\mathcal{K}).
\end{align}
Here, we used the fact that $L_\alpha$ has range on $t^{-1}(x)$.
As a result, it is possible to obtain a left-Hilbert algebra, and a von Neumann algebra, from the infinite qubit-chain groupoid $\mathcal G$ using the measure $\nu_\infty\times\#$.

\paragraph{Convolution.}
We now follow the step of Section~\ref{ch:groupoidalgebras} to obtain the left-Hilbert algebra $\mathfrak C$ and the von Neumann algebra of observables $\nu[\mathcal G]$ on the Hilbert space $\mathcal{H}=\mathrm{L}^2(\mathcal G,\mu)$.

Given a measurable function $F$ on $\mathcal{G}$, its Hahn norm is defined as in equation~\eqref{eq:hanhnorm_generalformula}, with $\Delta$ the modular function given in~\eqref{eq:modularfunction}, and the measures $\mu^x$ given by~\eqref{eq:measuredecomp_infinitecouplegroupoid}. The Hahn norm of a function $F$ is
\begin{equation}
	\label{eq:hanhnorminfinitequbitgroupoid}
	\lVert F\rVert_{\mathrm L^1_\mathrm H}=\max\left\{\left\lVert{\sum_{x^o\in\Gamma} \abs{F(x,x^o)}}\right\Vert_{\mathrm L^\infty},\left\lVert{\sum_{x^o\in\Gamma} \Delta(x,x^o)^{-1}\abs{F(x,x\oplus x^o)}}\right\Vert_{\mathrm L^\infty}\right\}
\end{equation} If $F,\ G$ are in $\mathrm L^2(\mathcal{G},\mu)$ and have finite Hahn norm, a convolution is well defined~\cite{lgcqt}. Let $\alpha=(x,x^o)$ in $\mathcal G$, and write the convolution of $F$ and $G$ as in~\eqref{eq:convolution_generalformula}
\begin{align}
	(F \star G) (\alpha)&= \int_{\mathcal G} F(\beta)G(\beta^{-1}\circ \alpha)d\mu^{t(\alpha)}(\beta)\nonumber \\
	&=\int_{\mathcal G}F(y,y^o)G((y\oplus y^o, y^o)\circ(x,x^o))d\mu^{x}(y,y^o)\nonumber\\
	&=\sum_{y^o\in\Gamma}F(x,y^o)G((x\oplus y^o,y^o)\circ(x,x^o))\nonumber\\
	&=\sum_{y^o\in\Gamma}F(x,y^o)G(x\oplus y^o,x^o\oplus y^o),
	\label{eq:convolution_infinitecouplegroupoid}
\end{align}
with $\beta=(y,y^o)$. The Dirac measure $\delta$ in~\eqref{eq:measuredecomp_infinitecouplegroupoid} ensures that $y=t(\beta)=t(\alpha)=x$. 

Call ${\mathrm L^1_\mathrm H(\mathcal G,\mu)}$ the space of function with finite Hahn norm, and $\mathfrak{A}$ the set of those who are also in $\mathrm L^2(\mathcal G,\mu)$:

\begin{equation}
	\label{eq:left_hilbert_algebra_qubit_groupoid}
	\mathfrak{C} = {\mathrm L^1_\mathrm H(\mathcal G,\mu)}\cap \mathrm L^2(\mathcal G,\mu).
\end{equation}
The convolution~\eqref{eq:measuredecomp_infinitecouplegroupoid} turns out to be associative on $\mathfrak C$, as the measure is $\sigma$-finite~\cite{lgcqt}.
The operator
\begin{align}\label{eq:regularrepresentation_infinitecouplegroupoid}
	\pi_F:\mathrm L^2(\mathcal G,\mu)&\rightarrow\mathrm L^2(\mathcal{G},\mu) \nonumber\\
	\psi&\mapsto \pi_F\psi=F\star\psi
\end{align}
is bounded. Indeed, given $F\in\mathfrak C$ and $\psi$ in $\mathrm L^2(\mathcal G)$, and using the Schwartz inequality, we get
\begin{align}\label{eq:inequalityconvolutionisbounded}
	\norm{F\star \psi}^2_{\mathrm L^2}&=\int_{\Omega_\infty}\sum_{x^o\in\Gamma}\Big|\sum_{y^o\in\Gamma}F(x,y^o)\psi(x\oplus y^o,x^o\oplus y^o)\Big|^2 d\nu_\infty(x)\nonumber\\
	&\leqslant \int_{\Omega_\infty}{\sum_{x^o\in\Gamma}\sum_{z^o\in\Gamma}\Big|{F(x,z^o)}\Big| \sum_{y^o\in\Gamma}\abs{F(x,y^o)\psi(x\oplus y^o,x^o\oplus y^o)^2}d\nu_\infty(x)}\nonumber\\
	&\leqslant \norm{F}_{\mathrm L_\mathrm H^1} \int_{\Omega_\infty}{\sum_{x^o\in\Gamma}  \sum_{y^o\in\Gamma}\abs{F(x\oplus y^o,y^o)\psi(x,x^o)^2} d\nu_\infty(x\oplus y^o)}\nonumber\\
	&=\norm{F}_{\mathrm L_\mathrm H^1} \int_{\Omega_\infty}\sum_{x^o\in\Gamma}|\psi(x,x^o)|^2\sum_{y^o\in\Gamma}|F(x\oplus y^o,y^o)|\Delta(x,y^o)^{-1}d\nu_\infty(x)\nonumber\\
	&\leqslant\norm{F}^2_{\mathrm L_\mathrm H^1}\norm{\psi}_{\mathrm L^2}^2,
\end{align}
so that $\norm{\pi_F}_{\mathcal B(\mathrm L^{2})}\leqslant\norm{F}_{\mathrm L_\mathrm H^1}$.
The vector space $\mathfrak C$ becomes a left-Hilbert algebra, with the involution $\dagger:\mathfrak C \rightarrow \mathfrak C$ in~\eqref{adjoint_generalformula} expressed as
\begin{equation}
	\label{eq:adjoint_algebra}
	F^{\dagger}(\alpha)=\Delta^{-1}(\alpha)\overline {F(\alpha^{-1})},
\end{equation}
and $\Delta$ given in equation~\eqref{eq:modularfunction}.
Finally, the Hilbert-algebra of operators can be closed with respect to the weak topology on $\mathrm L^2 (\mathcal G,\mu)$, thus obtaining the von Neumann algebra of observables.

From now on, we shall set all $\lambda_i$ equal to a given $\lambda$, with $0<\lambda\leqslant 1/2$. In the next section, we are going to show that this algebra of operators is isomorphic to the algebras defined by Powers in~\cite{Powers1967}, and we will do it by observing that they are both equivalent to the algebra defined by Puk\'{a}nszky in his work~\cite{pukanski_some}.

\subsection{Puk\'{a}nszky algebra}
We are now going to show that the physical algebra $\nu[\mathcal G]$ associated with the infinite qubit-chain groupoid is isomorphic to the algebra obtained by Puk\'{a}nszky~\cite{pukanski_some} and Powers~\cite{Powers1967}, and from this it will follow that different values of $\lambda$ yield different and nonequivalent quantum theories. In particular, for $0<\lambda<1/2$, we get factors of type $III$, while for $\lambda=1/2$ we get the hyperfinite type $II_1$ factor. In particular, the inequivalent type $III$ factors for $0<\lambda<1/2$ are indicated in literature as $\mathcal R_{\tilde\lambda}$, $\tilde\lambda=\lambda/(1-\lambda)\in[0,1]$~\cite{blackadar2006operator,connes1995noncommutative}.

Using the notation introduced above, the Puk\'{a}nszky algebra is defined on $\mathrm L^{2}(\mathcal{G},\mu)$, and it is generated by the two families of operators, $V_{x^o}, L_{\varphi}:\mathrm{L}^2(\mathcal G,\mu)\rightarrow \mathrm{L}^2(\mathcal G,\mu)$ given by
\begin{align}\label{eq:generatorialgebrapukanskiV}
	(V_{y^o}\psi)(x,x^o)&={\Delta(x,y^o)^{-1/2}}\psi(x\oplus y^o,x^o\oplus y^o),\\
	(L_{\varphi}\psi)(x,x^o)&=\varphi(x)\psi(x,x^o). \label{eq:generatorialgebrapukanskiL}
\end{align}
Here, $y^o$ is an element of $\Gamma$, while ${\varphi}$ is a measurable function in $\mathrm L ^{\infty}(\Omega_\infty, \nu_\infty)$. The term $\Delta^{-1/2}$ makes the operator $V_{y^o}$ unitary.

Let $\mathfrak M_\mathrm P$ be the von Neumann algebra generated by $V_{x^o}$ and ${L_{\varphi}}$.
We now prove that $\mathfrak M_\mathrm P=\nu[\mathcal{G}]$. In order to do this, we just need to prove that the generating operators of both algebras belong to each others.

We start with $\mathfrak M_\mathrm P\subset \nu[\mathcal G]$. Let $y^o$ be in $\Gamma$, and we look for a function $F_{y^o}\in \mathrm L^2\cap \mathrm L^1_\mathrm H$ such that
\begin{equation}
	(V_{y^o}\psi)(x,x^o)=(F_{y^o}\star \psi) (x,x^o)=\sum_{z^o\in\Gamma}F_{y^o}(x,z^o)\psi(x\oplus z^o,x^o\oplus z^o),
\end{equation}
for all $\psi\in\mathrm L^2$. This implies $F_{y^o}(x,x^o)=\delta_{y^o}(x^o)\Delta(x,x^o)^{-{1}/{2}}$. Moreover, $F_{y^o}$ is indeed in $\mathrm L^2\cap \mathrm L^1_\mathrm H$. Explicitly, $\left\lVert F_{y^o}\right\lVert_{\mathrm L^2}=1$, while
\begin{align}
	&\sum_{x^o\in\Gamma}\abs{F_{y^o}(x,x^o)}={\Delta(x,y^o)}^{-1/2}\leqslant \left(\frac{1-\lambda}{\lambda}\right)^{\sum_{j\in\mathbb N}y^o_j/2}<+\infty,\\
	&\sum_{x^o\in\Gamma} \abs{F_{y^o}(x,x^o)}\Delta^{-1}(x,x^o)={\Delta(x,y^o)}^{-3/2}\leqslant \left(\frac{1-\lambda}{\lambda}\right)^{\sum_{j\in\mathbb N}3y^o_j/2}<+\infty.
\end{align}
So, $V_{y^o}\in\nu[\mathcal G]$. Regarding~\eqref{eq:generatorialgebrapukanskiL}, let $\varphi\in\mathrm L^\infty(\Omega_\infty)$, and consider the equation
\begin{equation}
	(L_\varphi\psi)(x,x^o)=(G_\varphi\star\psi)(x,x^o)=\sum_{z^o\in\Gamma}G_\varphi(x,z^o)\psi(x\oplus z^o,x^o\oplus z^o).
\end{equation}
Then, we get
\begin{equation}
	G_\varphi(x,x^o)=\delta_{\{0\}}(x^o)\varphi(x).
\end{equation}
It is easy to see that $\norm{G_\varphi}_{\mathrm L^2}\leqslant\norm{\varphi}_{\mathrm L^\infty}$, while 
\begin{align}
	&\left\Vert \sum_{x^o\in\Gamma}\abs{G_\varphi(x,x^o)} \right\Vert_{\mathrm L^\infty}=\left\Vert \sum_{x^o\in\Gamma}\delta_{\{0\}}(x^o)\abs{\varphi(x)} \right\Vert_{\mathrm L^\infty}=\left\Vert \varphi \right\Vert_{\mathrm L^\infty} <+\infty,\\
	&\left\Vert \sum_{x^o\in\Gamma} \abs{G_\varphi(x,x^o)} \Delta(x,x^o)^{-1} \right\Vert_{\mathrm L^\infty}= \left\Vert \sum_{x^o\in\Gamma} \delta_{\{0\}}(x^o)\abs{\varphi(x)}\Delta(x,x^o)^{-1}\right\Vert_{\mathrm L^\infty}=\left\Vert \varphi \right\Vert_{\mathrm L^\infty} \nonumber\\&\qquad\qquad\qquad\qquad\qquad\qquad\qquad<+\infty,
\end{align}
so that $G_\varphi\in\mathfrak C$, and $L_\varphi\in\nu[\mathcal G]$. In particular, note that the operators $L_\varphi$ define an Abelian subalgebra which we can be interpreted as the classical algebra corresponding to a noiseless measurement on the spins. As a result, $\mathfrak M_\mathrm P\subset \nu[\mathcal G]$.

Conversely, we should prove that a function in $\mathfrak C$ is in the Puk\'{a}nszky algebra. First, observe that for all $y^o\in\Gamma$ and $\varphi\in\mathrm L^\infty(\Omega_\infty)$, the function 
\begin{equation}\label{eq:functionwithacipointasureongamma}
	F(x,x^o)=\delta_{\{y^o\}}(x^o)\varphi(x)
\end{equation}
is in $\mathrm L^2\cap\mathrm L^1_{\mathrm H}$, and also
\begin{align}
	(F\star \psi)(x,x^o)
	&=\sum_{z^o\in\Gamma}\delta_{\{y^o\}}(z^o)\varphi(x)\psi(x\oplus z^o,x^o\oplus z^o)=\varphi(x) \psi(x\oplus z^o,x^o\oplus z^o)\nonumber\\
	&=\varphi(x){\Delta(x,y^o)}^{1/2}(V_{y^o}\psi)(x,x^o)=L_{\varphi\sqrt\Delta}V_{y^o}\psi(x,x^o).
\end{align}
Therefore, a function in the form~\eqref{eq:functionwithacipointasureongamma} defines an operator $\pi_F$~\eqref{eq:regularrepresentation_infinitecouplegroupoid} via the convolution~\eqref{eq:convolution_infinitecouplegroupoid}, and this operator is in the Puk\'{a}nszky algebra $\mathfrak M_\mathrm P$. Next, we decompose any other function $F$ in $\mathfrak{C}$ as a sum of functions of type~\eqref{eq:functionwithacipointasureongamma}. Explicitly
\begin{equation}
	F(x,x^o)=\sum_{y^o\in\Gamma}\delta_{\{y^o\}}(x^o)F(x,x^o)\,,
\end{equation}
where $\left\Vert F(x,x^o) \right\Vert_{\mathrm{L}_{\infty}} \leq \norm{F}_{\mathrm L_\mathrm H^1} < \infty$. This means that every element in the previous sum is in the form~\eqref{eq:functionwithacipointasureongamma} and thus it is in the Puk\'{a}nszky algebra. Consequently, any finite sum of such elements is in the algebra, too. In order to prove that $F$ is in the algebra, we just need to prove that
\begin{equation}
	(\chi_{\Gamma_n}F)\star \psi\xrightarrow{\mathrm L^2} F \psi,
\end{equation}
with $\Gamma_n=\{x^o\in\Gamma|x_i=0,\ \forall i>n\}$, and $\chi_{\Gamma_n}$ the characteristic function over $\Gamma_n$. By calling $G_n=F(1-\chi_{\Gamma_n})$ and repeating the same steps as in~\eqref{eq:inequalityconvolutionisbounded}, we have
\begin{align}
	\norm{G_n\star\psi}_{\mathrm L^2}^2 &= \int_{\Omega_\infty} \sum_{x^o\in\Gamma}\Big|\sum_{y^o\in\Gamma}G_n(x,y^o)\psi(x\oplus y^o,x^o\oplus y^o)\Big|^2d\nu_\infty(x)\nonumber\\
	&\leqslant \int_{\Omega_\infty}{\sum_{x^oz^o\in\Gamma}\Big|{G_n(x,z^o)}\Big| \sum_{y^o\in\Gamma}\abs{G_n(x,y^o)\psi(x\oplus y^o,x^o\oplus y^o)^2} d\nu_\infty(x)}\nonumber\\
	&\leqslant \norm{F}_{\mathrm L^1_H}\sum_{y^o\notin \Gamma_n}\left\{\int_{\Omega_\infty}{\sum_{x^o\in\Gamma} \abs{F(x,y^o)\psi(x\oplus y^o,x^o\oplus y^o)^2} d\nu_\infty(x)}\right\}.
\end{align}
The last element in the equation converges to $0$ if the sum over all $y^o$ is convergent. But this is the case as
\begin{equation}
	\sum_{y^o\in \Gamma_n}\left\{\int_{\Omega_\infty}{\sum_{x^o\in\Gamma} \abs{F(x,y^o)\psi(x\oplus y^o,x^o\oplus y^o)^2} d\nu_\infty(x)}\right\}\leqslant \norm{F}_{\mathrm L^1_H}\norm{\psi}_{\mathrm L^2}^2<\infty,
\end{equation}
by repeating the same computation of the last line in equation~\eqref{eq:inequalityconvolutionisbounded}. As a result, $\pi_F\in\mathfrak M_\mathrm P$, and $\mathfrak M_\mathrm P\supset\nu[\mathcal G]$, so that $\mathfrak M_\mathrm P=\nu[\mathcal G]$. 

In the final part of this section we will show how one can obtain the Puk\'{a}nszky algebra $\mathfrak M_\mathrm P$ from the Powers construction given in Section~\ref{sec:C-W algebras}. In particular, we will represent the algebra $\mathfrak A$ in equation~\eqref{eq:powercstaralgebra} on the Hilbert space $\mathrm L^2(\mathcal G, \mu)$, obtaining an equivalent cyclic representation. This construction was first obtained by Glimm~\cite{glimms61}, and maps the Puk\'{a}nszky factors generated by~\eqref{eq:generatorialgebrapukanskiV} and~\eqref{eq:generatorialgebrapukanskiL} to the ones obtained by Powers. In order to get the corresponding cyclic representation, it is necessary for the parameter $\lambda$ in the measure on $\Omega$ of Eq.~\eqref{eq:singlesiteprobability} to be the same $\lambda$ as in~\eqref{eq:powerstateproduct}.

For $k\in\mathbb N$, consider $V_{e_k}$ and $L_{\psi_k}$, with  
\begin{align}\label{eq:orthogonalbasisinfinitecase}
	e_k&=(\delta_{kj})_{j\in\mathbb N}=(0,\dots,0,\overset{\substack{\text{k} \\ \downarrow }}{1},0,\dots)\in\Gamma,\\
	\label{eq:orthogonalbasisinfinitecaseII}
	\psi_k(x)&=
	\begin{cases}
		+1\ \text{if}\ x_k=0,\\
		-1\ \text{if}\ x_k=1.
	\end{cases}\in\mathrm L^{\infty}(\mathcal G, \mu).
\end{align}
The Glimm representation $\pi_\lambda$ of $\mathfrak{A}$ on $\mathrm{L}^2 (\mathcal G,\mu)$ is defined by assigning
\begin{align}
	\label{eq:sigmazk}
	\sigma_1^{(k)}=\underbrace{\mathbb I_2 \otimes \mathbb I_2\otimes\dots\otimes\mathbb I_2}_{k-1}\otimes \sigma_1\in M_{2^k}\mapsto V_{e_k}\in\nu_\mathrm P,\\
	\sigma_3^{(k)}=\underbrace{\mathbb I_2 \otimes \mathbb I_2\otimes\dots\otimes\mathbb I_2}_{k-1}\otimes \sigma_3\in M_{2^k}\mapsto L_{\psi_k}\in\nu_\mathrm P.
\end{align}
The cyclic vector is
\begin{equation}
	\Psi_\lambda(x,x^o)=\delta_{\{0\}}(x^o)\chi_{\Omega_{\infty}}(x),
\end{equation}
and $\braket{\Psi_\lambda}{\pi_\lambda(A)\Psi_\lambda}=\phi(A)$. To prove this, just expand an $A$ in $M_{2^n}$ in a tensor product basis of $\{\mathbb I_2,\sigma_1,\sigma_2,\sigma_3\}$ and evaluate explicitly the expectation values of $\Psi_\lambda
$.

\subsection{Modular theory}

Given a left-Hilbert algebra, it is possible to write the polar decomposition of the closure of the involution $\dagger$, as in equation~\eqref{eq:TTpolardecomposition}. We will indicate with $\hat\Delta$ the modular operator obtained from Tomita-Takesaki, to distinguish it from the modular function $\Delta$ in~\eqref{eq:modularfunction}.

As we have already mentioned in Section~\ref{ch:groupoidalgebras} the modular operator and the modular conjugation are connected to the modular function $\Delta$ of the groupoid $\mathcal G$ by
\begin{align}
	(JF)(\alpha)&= \Delta^{-1/2}(\alpha)\overline F(\alpha^{-1}),\\
	(\hat\Delta F)(\alpha)&= \Delta(\alpha) F(\alpha),
\end{align}
for $F$ in $\mathfrak C$, see~\cite{lgcqt}.
It turns out that $\Delta$ is just a multiplication operator. Explicitly, by using~\eqref{eq:modularfunction}, one gets
\begin{equation}
	\hat\Delta(x,x^o)=\prod_{j=1}^\infty\left(\frac{1-\lambda}{\lambda}\right)^{(2x_j-1)x^o_j}.
\end{equation}
From Tomita-Takesaki modular theory, and in particular from Theorem~\ref{th:TTtheorem}, it is possible to define a unitary dynamics on the algebra $\nu[\mathcal G]$, see equation~\eqref{eq:modulardynamics}. The Hamiltonian generating this dynamics is given by the logarithm of the Modular operator $\hat\Delta$,  which in our case explicitly reads
\begin{equation}
	\label{eq:HamiltonianTTmodulartheory}
	H(x,x^o)=\log\hat\Delta(x,x^o)=\log\frac{1-\lambda}{\lambda}\sum_{j=1}^\infty (2x_j-1)x^o_j.
\end{equation}

The corresponding spectrum is the essential range of the multiplication function, namely:
\begin{equation}
	\mathrm{spec}(H)=\log\left(\frac{1-\lambda}{\lambda}\right) k,\qquad k\in\mathbb Z.
\end{equation}
Explicitly, for each site $j$, each term in the sum takes the value $+1$ for $x_j^0=1$ and $x_j=1$, while it is $-1$ for $x_j^0=1$ and $x_j=0$. The corresponding eigenvectors are obtained as combination of characteristic function over these sets. Compare this spectrum with~\cite[p.~485]{connes1995noncommutative}.

\section{DFS states}
\label{sec:DFS_states}
In the groupoid picture, there are different ways to introduce states. From one point of view, one could define a state as a quantum measure on the groupoid, where a quantum measure is a generalization of measure theory introduced by Sorkin to take into account interference~\cite{sorking}. On the other hand, once the groupoid algebra is defined, states can be described as positive normalized functionals over the algebra of observables $\nu[\mathcal G]$, inheriting the structure of the algebraic description~\cite{groupoidII}. In this paper we will follow this second approach. 

Between states defined in the groupoid picture, a family which turns out to be particularly relevant is the family of Dirac-Feynman-Schwinger  states~\cite{ciagilapropagator}. For the groupoid of pairs over a measurable space $\Omega$ (see Example~\ref{example:standardcouplegroupoid}), they turn out to be the pure states of the theory, once the physical algebra is introduced (Eq.~\eqref{eq:standardalgebra}). However, their major interest is related to the possibility of encoding the dynamics of a system in the histories approach to quantum mechanics~\cite{ciagilapropagator,ciagliacausality,Sorkin_2011}. Indeed, these states are represented via a function of positive-type on the groupoid which is also a homomorphism of the groupoid with values in $U(1)$, and the corresponding phase factor can be interpreted as an abstract action functional for the groupoid of paths of a kinematical groupoid.

On the other hand, DFS functions can be used to define homomorphisms of the groupoid with values in $\mathbb{R}_+$, if one replaces the imaginary exponential with a real exponential. We have already seen an example of such function, the function $\Delta$ of the infinite qubit-chain which has been proved to be associated with the Tomita-Takesaki operator of the corresponding groupoid-algebra. In this sense, DFS functions can determine dynamics via the automorphism of the groupoid-algebra associated with the Tomita-Takesaki operator. In this paper, we will look at DFS functions from this perspective and we will firstly arrive at a complete characterization of them, using the fact that they are connected to a certain cohomology class of the group $\Gamma$. Then, one can ask oneself if there is a suitable Haar system of measures on the groupoid which has the chosen real DFS function as modular function. The answer is affirmative as one can see in~\cite{Connes1979,Kastler1982} and we will provide an example of interest connected with Ising dynamics. Let us start with the first result.

For the infinite qubit-chain, a real DFS state is defined starting from a function of positive type of the form
\begin{equation}\label{eq:generalschwingerstate}
	\varphi(\alpha)=\mathrm e^{- \mathscr S(\alpha)}.
\end{equation}
where $\mathscr S:\mathcal G\rightarrow \mathbb R$ is a measurable function which satisfies the condition
\begin{equation}\label{eq:dfscompositioncondition_generalformula}
	\mathscr S(\alpha\circ \beta)=\mathscr S(\alpha)+\mathscr S(\beta).
\end{equation}

We will now give a constructive procedure to obtain real DFS function for the infinite qubit-chain groupoid.

The homomorphism condition~\eqref{eq:dfscompositioncondition_generalformula} reads
\begin{equation}\label{eq:proprietalogaritmicadischwingerperilgruppoidedellecoppieinfinite}
	\mathscr S(x,x^o\oplus y^o)=\mathscr S(x\oplus y^o,x^o)+  \mathscr S(x,y^o)= \mathscr S(x,x^o)+\mathscr S(x\oplus x^o,y^o).
\end{equation}
In a first approach, one could simply define $\mathscr S$ on the elements $(x,e_{k})$, with $\{e_k\}_{k\in\mathbb N}$ being the elements in $\Gamma$ in equation~\eqref{eq:baseinfinitecouplegroupoid}, and $x$ in $\Omega_\infty$. Then, this expression should be extended to all $(x,x^o)$ in $\Omega_\infty\times\Gamma$ by writing $x^o=\sum_i e_{k_i}$ and using the first equality of~\eqref{eq:proprietalogaritmicadischwingerperilgruppoidedellecoppieinfinite}. Unfortunately, the second equality introduces other constraints on the space of DFS functions. Note that it is a consequence of the commutativity of the group $\Gamma$.

We can refine the previous idea, and use it to construct a DFS function inductively.
First, we need to observe some basic properties of a DFS function. In particular, $\mathscr S$ has to satisfy
\begin{equation}
	\mathscr S(x,0)=0,
\end{equation}
as $\mathscr S(x,x^o)=\mathscr S(x,x^o\oplus0 )=\mathscr S(x,0)+\mathscr S(x,x^o)$. Moreover
\begin{equation}\label{eq:inversionofschwinger}
	\mathscr 	S(x\oplus x^o,x^o)=\mathscr S(x,0)-\mathscr S(x,x^o)=-\mathscr S(x,x^o),
\end{equation}
We will use these two properties to inductively construct $\mathscr S$. Define for $n\in\mathbb N$ the sets
\begin{align}
	\Gamma_n&=\{(x_1,x_2\dots)\in\Omega_\infty: x_k=0,\ \forall k>n \},\\
	\mathscr C_n&=\{(x_1,x_2\dots)\in\Omega_\infty: x_k=0,\ \forall k\leqslant n \}.
\end{align}
Note that $\Gamma_n\subset \Gamma=\cup_k \Gamma_k$, while $\mathscr C_n \subset\Omega_\infty$ is the cylinder $\mathscr C(\{\underbrace{0,0,\dots,0}_{n}\})$, see equation~\eqref{eq:definition_cylinder}.

We will use induction hypothesis by requiring that $\mathscr S$ satisfies the DFS condition on the set
\begin{equation}
	\Omega_\infty\times\Gamma_n,
\end{equation}
which means that~\eqref{eq:proprietalogaritmicadischwingerperilgruppoidedellecoppieinfinite} is satisfied for all $x^o,y^o\in\Gamma_{n}$ and $x\in\Omega_\infty$. Then, we will extend $\mathscr S$ to a function satisfying DFS condition on $\Omega_\infty\times \Gamma_{n+1}$. The process is performed in the following way.
\begin{enumerate}
	\item[(1)] Define freely the function $\mathscr S(x,e_{n+1})$ for $x$ in the cylinder $\mathscr C_{n+1}$. This means that $x_1=x_2=\dots=x_{n+1}=0$, and $\mathscr S$ is defined on $\mathscr{C}_{n+1}\times\{e_{n+1}\}$.
	\item[(2)] Extend it on $\mathscr C_n$ using condition~\eqref{eq:inversionofschwinger}. This means defining $\mathscr S(x\oplus e_{n+1},e_{n+1})\equiv-\mathscr S(x,e_{n+1})$, with $x$ again in $\mathscr C_{n+1}$. At this point, $\mathscr S$ is defined on $\Omega_\infty\times \Gamma_n$ and $\mathscr{C}_n\times\{e_{n+1}\}$.
	\item[(3)] Consider $x^o\in\Gamma_n$ and $z\in\Omega_\infty$. Decompose uniquely $z = \overline z\oplus z^o$, with $z^o\in\Gamma_n$ and $\overline z$ in $\mathscr C_n$. Thus, define
	\begin{equation}\label{eq:formulatoextenddfs}
		\mathscr S(z,x^o\oplus e_{n+1})\equiv\mathscr S\underbrace{(\overline z \oplus e_{n+1},z^o\oplus x^o)}_{\in\ \Omega_\infty\times \Gamma_n}-\underbrace{\mathscr S(\overline z ,z^o)}_{\in\Omega_\infty\times \Gamma_n} + \underbrace{\mathscr S(\overline z,e_{n+1})}_{\in\mathscr{C}_n\times \{e_{n+1}\}}.
	\end{equation}
\end{enumerate}
Since $\Gamma_{n+1}=\Gamma_n\cup(\Gamma_n\oplus \{e_{n+1}\})$, equation~\eqref{eq:formulatoextenddfs} extends $\mathscr S$ on the whole $\Omega_\infty\times\Gamma_{n+1}$.
We must prove that it is indeed a DFS function on $\Omega_\infty\times \Gamma_{n+1}$. Explicitly, we need to show that
\begin{equation}\label{eq:dfsconditioninfinitecouplegroupoid}
	\mathscr S(z,x^o\oplus y^o)= \mathscr S(z\oplus y^o,x^o)+\mathscr S(z,y^o)=\mathscr S(z,x^o)+\mathscr S(z\oplus x^o,y^o)
\end{equation}
for all $z\in\Omega_\infty$ and $x^o,y^o\in\Gamma_{n+1}$. We will check it by evaluating each term for all	 possible values of $x^o_{n+1}$ and $y^o_{n+1}$.\\

\textbf{Case 1: $x_{n+1}^o=1,\ y_{n+1}^o=0$.}

Recall $x^o=e_{n+1}\oplus x^o$, so that $x^o\in \Gamma_n$. Decompose $z=z^o\oplus\overline z$, with $z^o\in\Gamma_n$ and $\overline z\in\mathscr C_n$. We check the first equality in~\eqref{eq:dfsconditioninfinitecouplegroupoid}.
\begin{align}
	\mathscr S(\overline z\oplus z^o,(x^o\oplus e_{n+1})\oplus y^o)& =\mathscr S( \overline z \oplus e_{n+1},x^o\oplus y^o \oplus z^o)-\mathscr S(\overline z,z^o) +\mathscr S(\overline z,e_{n+1})\nonumber\\
	&=\mathscr S( \overline z \oplus e_{n+1},x^o\oplus (y^o \oplus z^o))-\mathscr S(\overline z, y^o\oplus z^o) \nonumber\\
	&\quad+\mathscr S(\overline z,e_{n+1})+\mathscr S(\overline z, y^o\oplus z^o)-\mathscr S(\overline z,z^o)\nonumber\\
	&=	\mathscr S(\overline z\oplus y^o \oplus z^o,x^o+e_{n+1})+ \mathscr S(\overline z\oplus z^o,y^o)\nonumber \\
	&= \mathscr S(z\oplus y^o,x^o\oplus e_{n+1})+\mathscr S(z,y^o)
\end{align}
For the second equality in~\eqref{eq:dfsconditioninfinitecouplegroupoid}
\begin{align}
	\mathscr S(\overline z\oplus z^o,(x^o\oplus e_{n+1})\oplus y^o)& =\mathscr S( \overline z \oplus e_{n+1},x^o\oplus y^o \oplus z^o)-\mathscr S(\overline z,z^o) +\mathscr S(\overline z,e_{n+1})\nonumber\\
	&=\mathscr S( \overline z \oplus e_{n+1},x^o\oplus z^o)-\mathscr S(\overline z,z^o)+\mathscr S(\overline z,e_{n+1})\nonumber\\
	&\quad+S( \overline z\oplus z^o \oplus x^o\oplus e_{n+1} ,y^o)\nonumber \\
	&= \mathscr S(z,x^o\oplus e_{n+1})+\mathscr S(z\oplus x^o\oplus e_{n+1},y^o).
\end{align}
This proves equation~\eqref{eq:dfsconditioninfinitecouplegroupoid} in the first case.\\

\textbf{Case 2:} $x_{n+1}^o=y_{n+1}^o=1$.

Again, recall $x^o= x^o\oplus e_{n+1}$, $y^o=y^o\oplus e_{n+1}$, so that $x^o,y^o\in\Gamma_n$. Note that $S(z,(x^o \oplus e_{n+1}) \oplus (y^o \oplus e_{n+1}))= S(z,x^o\oplus  y^o)$. On the other side
\begin{align}
	\mathscr S(z\oplus (y^o\oplus e_{n+1}),x^o\oplus e_{n+1})&=\mathscr S((\overline z \oplus e_{n+1})\oplus (y^o\oplus z^o),x^o \oplus e_{n+1})\nonumber \\
	&=\mathscr S((\overline z\oplus e_{n+1})\oplus e_{n+1}, x^o\oplus y^o \oplus z^o)\nonumber \\
	&\quad -\mathscr S(\overline z\oplus e_{n+1},y^o\oplus z^o)
	+\mathscr S (\overline z\oplus e_{n+1},e_{n+1})\nonumber\\
	&=\mathscr S(\overline z, x^o\oplus y^o \oplus z^o)-\mathscr S(\overline z\oplus e_{n+1},y^o\oplus z^o)\nonumber\\
	&\quad-\mathscr S(\overline z,e_{n+1}).
\end{align}
As a result
\begin{align}
	\mathscr S(z\oplus (y^o\oplus e_{n+1})x^o\oplus e_{n+1})&+\mathscr S(z,y^o\oplus e_{n+1})= \mathscr S(\overline z, x^o\oplus y^o \oplus z^o)\nonumber\\
	&\quad-\mathscr S(\overline z\oplus e_{n+1},y^o\oplus z^o)-\mathscr S(\overline z,e_{n+1})\nonumber\\
	&\quad+\mathscr S(\overline z\oplus e_{n+1},z^o\oplus y^o)-\mathscr S(\overline z,z)-\mathscr S(\overline z,z^o)\nonumber\\
	&=\mathscr S(z\oplus z^o,x^o\oplus y^o).
\end{align}
In this way, we see that equation~\eqref{eq:formulatoextenddfs} generates a the real DFS function on $\Omega_\infty\times \Gamma_n$. By repeating this procedure, it is possible to extend it to the whole space $\Omega_\infty\times \Gamma$. 

Let us note here that the condition \eqref{eq:proprietalogaritmicadischwingerperilgruppoidedellecoppieinfinite} can be interpreted also in terms of a cohomology of the group $\Gamma$. Indeed, let $\mathscr A$ be the space of measurable functions on $\Omega_{\infty}$ which is a $\Gamma$-module~\cite{Moore1964} under the following action of the group $\Gamma$:
\begin{equation}
	(x^o \circ S) (x) = S(x\oplus x^o)\,.
\end{equation}
In particular, observe that a $\Gamma$-module can be interpreted as a function from $\Omega_\infty\times \Gamma$ in $\mathbb R$.

Call $\Gamma^n$ the direct product of $n$ copies of $\Gamma$
\begin{equation}
	\Gamma^n= \underbrace{\Gamma\times \Gamma \times \dots\times \Gamma}_{n}.
\end{equation}
The cochains of order $n$ of this cohomology are the functions from $\Gamma^n$ to $\mathscr A$, defined recursively via the differential operator $\delta^n \,\colon\, C_n\,\rightarrow \, C_{n+1}$ in the following way
\begin{align}
	(\delta^n c) (x^o_1,x^o_2,\cdots,x^o_{n+1}) &= (x^o_1 \circ c) (x^o_2,\cdots,x^o_{n+1}) + \nonumber\\
	\sum_{k=1}^{n}(-1)^k &c(x^o_1,x^o_2,\cdots, x^o_k\oplus x^o_{k+1}, \cdots,x^o_{n+1}) + (-1)^{n+1} c(x^o_1,x^o_2,\cdots,x^o_n)\,. 
\end{align}
In the previous formula, $C_n$ is the set of cochains of order $n$, and $c$ is a function in $C_n$. In particular, DFS functions will satisfy the following 2-cocycle condition:
\begin{equation}
	(\delta^1 S)(x^o, y^o)[x] = (x^o\circ S(y^o))[x] - S(x^o\oplus y^o)[x] + S(x^o)[x] = 0 .
\end{equation} 
Therefore, as a special class of solutions of this 2-cocycle condition, there are cochains which are exact, which in this case satisfies the following condition:
\begin{equation}
	S(x^o)[x] = (\delta^0 H)(x^o)[x] = (x^o \circ H)[x] - H[x] 
\end{equation}
which are expressed only in terms of a measurable function $H\,\colon\,\Omega_{\infty}\,\rightarrow\,\mathbb{R}$. 
In the next section we are going to present an example where we use one of these exact cochains as a modular function of a measured groupoid, and the associated Tomita-Takesaki dynamics can be interpreted in terms of the well-known Ising model.

\subsection{The Ising model}
In this section, we are going to use the infinite qubit chain groupoid~\eqref{eq:infinitequbitchaingroupoid} to describe the Ising model for an infinite chain. It is a standard problem in thermodynamics. The system can be solved in the finite dimensional case by considering the Jordan-Wigner transformation~\cite{wigner1928paulische,Lieb1961,invernizzi}. Nevertheless, the problem for the infinite dimensional chain is well defined only for proper boundary conditions.
By using the groupoid formalism it is possible to obtain a well defined Hamiltonian by defining the energy corresponding to a transition, that is, how the total energy changes for the flip of a spin, despite the fact that the total energy of an infinite chain diverges. 
The Ising model for a chain of spins at zero magnetic field is written as
\begin{equation}
	H=-J\sum_{k} \sigma_3^{(k)}\sigma_3^{(k+1)}.
\end{equation}
For simplicity, we will consider the $z$ axis as reference, so that it will be easier to represent the operator on the groupoid chain~\eqref{eq:infinitequbitchaingroupoid}. Moreover, we use the same convention as in equation~\eqref{eq:sigmazk}. Using the Glimm's map~\eqref{eq:sigmazk}, we can explicitly write  the action of $H$ on the groupoid space $\mathrm L^2(\mathcal G,\mu)$:
\begin{equation}
	\label{eq:ising_model_hamiltonian}
	H(x)=-\sum_{k\in\mathbb N} J\psi_k(x)\psi_{k+1}(x),
\end{equation}
with $\psi_k(x)$ defined in equation~\eqref{eq:orthogonalbasisinfinitecaseII}. Now,  observe that in general this Hamiltonian is not well defined, as the series here is not absolutely convergent, since $\abs*{\psi_k(x)}=1$.
However, by using the groupoid formalism, we can actually obtain a finite and well defined Hamiltonian which gives the change in energy corresponding to a transition.
It is formally obtained from $H$ as
\begin{equation}
	\label{eq:differential_hamiltonianold}
	S(x,x^o)=	H(x\oplus x^o)-H(x).
\end{equation}
Since the two functions are not well defined separately, we give the explicit expression of it as
\begin{equation}
	\label{eq:differential_hamiltonian}
	S(x,x^o)= -J\sum_{k\in\mathbb N} (\psi_k(x\oplus x^o)\psi_{k+1}(x\oplus x^o)-\psi_k(x)\psi_{k+1}(x)).
\end{equation}

The function $S$ is now well defined as for all $x$ and $x^o$ the sum is convergent, as $k$ now runs only over a finite number of terms. Explicitly, this is a multiplication self-adjoint operator on $\mathrm L^2(\mathcal G, \mu)$, and (up to a sign) it is nothing but the difference in energy between the source and the target of the transition $(x,x^o)$. We can evaluate it, for example, for $x^o=e_k$, as defined in~\eqref{eq:orthogonalbasisinfinitecase}. Explicitly, one gets
\begin{align}
	S(x,e_k)&=-J(\psi_k(x\oplus e_k) \psi_{k+1}(x) - \psi_k(x)\psi_{k+1}(x) +\psi_{k-1}(x)\psi_{k}(x\oplus e_k)\nonumber\\ &\quad -\psi_{k-1}(x)\psi_{k}(x))\nonumber\\
	&= -J\psi_{k+1}(x)(\psi_k(x\oplus e_k)-\psi_k(x)) -J\psi_{k-1}(x)(\psi_{k}(x\oplus e_k)-\psi_{k}(x))\nonumber\\
	&= -J(\psi_{k+1}(x)+\psi_{k-1}(x))(\psi_{k}(x\oplus e_k)-\psi_{k}(x)).
\end{align}
In other words, the flip of a spin will change the energy of the whole system depending of the sign of its neighbors spins.

The issue that we are going to address now is the following: can we find a measure on the groupoid $\mathcal G$ such that the function
\begin{equation}
	\label{eq:modular_function_easing}
	\Delta_H(x,x^o) = \mathrm{e}^{-S(x,x^o)}.
\end{equation}
will be the associated modular function? The answer is affirmative. Indeed, in Connes' notation, a Haar system of measures on the groupoid is nothing but a transverse function~\cite{Connes1979,Kastler1982}. Once a modular function is chosen, which is a homomorphism of the groupoid $\Delta\,\colon\,\mathcal{G}\,\rightarrow \,\mathbb{R}_+$ with values in the group of positive real numbers, every transverse measure provides a measure on the base space of the groupoid such that we have a desintegrated measure on the whole groupoid absolutely continuous with respect to its inverse. The corresponding modular function will be the chosen $\Delta$. A transverse measure in Connes' noncommutative integration is a functional $\Lambda\,\colon\,\mathcal{E}_+\,\rightarrow\,\overline{\mathbb{R}}_+$ from the space of transverse functions to the extended positive real numbers satisfying certain conditions. Introducing them properly would lead us out of the scope of this work and we refer to the cited works for  the details. We have mentioned it to show that the issue we are posing at this moment can be addressed in a general framework using Connes' noncommutative integration theory. However, in the rest of this section we are going to explicitly construct a measure on the space $\Omega_{\infty}$ such that the associated modular function is indeed $\Delta_H$. 

In order to build this measure we will introduce a family of probability spaces $\left\lbrace (\Omega_{\times n}, \nu_H^{(n)}) \right\rbrace_{n\in\mathbb{N}}$, and, as in the previous analysis, we will make use of the Kolmogorov theorem to define a probability $\nu_H$ 
on $\Omega_\infty$.
By calling
\begin{equation}
	\Omega_{\times n} = \prod_{i=1}^{n} \Omega_i
\end{equation}
we approximate the Ising Hamiltonian~\eqref{eq:ising_model_hamiltonian} on $\Omega_{\times n}$ as follows:
\begin{equation}
	H_n(x^{(n)})=-J\sum_{k=1}^{n-1} \psi_k(x)\psi_{k+1}(x)=-J\sum_{k=1}^{n-1}(2x_k-1)(2x_{k+1}-1),
\end{equation}
with $x^{(n)}=(x_1,\cdots,\,x_n)$ an element in $\Omega_{\times n}$.
Note the free condition on the boundary. We then consider the probability on $\Omega_{\times n}$ defined via the Boltzmann distribution function
\begin{equation}
	\nu_H^{(n)}(A_n) = \frac{1}{Z_n}\sum_{x^{(n)} \in A_n} \mathrm{e}^{H_n(x)},
\end{equation}
with $A_n$ a generic subset of $\Omega_{\times n}$, and $x^{(n)}=(x_1,\cdots,\,x_n)$. The normalization factor $Z_n$ is the partition function
\begin{align}
	Z_n&=\sum_{x^{(n)}\in \Omega_{\times n}} \mathrm{e}^{H_n(x^{(n)})} =\sum_{\bar x_1,\dots,\bar x_n=\pm 1} \mathrm e^{-J(\bar x_1\bar x_2+\bar x_2\bar x_3+\dots+ \bar x_{n-1}\bar x_n)}\nonumber\\
	&= \sum_{\bar x_1,\dots,\bar x_{n-1}=\pm 1} \mathrm e^{-J(\bar x_1\bar x_2+\bar x_2\bar x_3+\dots+ \bar x_{n-2}\bar x_{n-1})} 2 \cosh(J\bar x_{n-1})\nonumber\\
	&=2 \cosh J \sum_{\bar x_1,\dots,\bar x_{n-1}=\pm 1}\mathrm e^{-J(\bar x_1\bar x_2+\bar x_2\bar x_3+\dots+ \bar x_{n-2}\bar x_{n-1})}= (2 \cosh J)^n.
\end{align}
Here, for simplicity, we called $\bar x_k=2x_k-1$. Note that in the second line the function $\cosh (J \bar x_n)$ is even, and it is independent of $x_n$.

Consider the family of surjective maps $\pi_{n,k}\,\colon\,\Omega_{\times n}\,\rightarrow\,\Omega_{\times k}$, defined for $n>k$, which acts as $\pi_{n,k}(x_1,x_2,\cdots,x_n) = (x_1,x_2,\cdots,x_k)$, so that $\pi_{n,k}\circ \pi_{k,j} = \pi_{n,j}$ ($n>k>j$). A straightforward application of the definition of push-forward of measures shows that $(\pi_{n,k})_{\ast}\nu_H^{(n)}=\nu_H^{(k)}$. Indeed, call $A_k$ a subset of $\Omega_{\times k}$, and consider:
\begin{align}
	(\pi_{n,k})_{\ast}\nu_H^{(n)}(A_k)&=\nu_H^{(n)}(\pi_{n,k}^{-1}(A_k))=\nu_H^{(n)}(A_k\times \Omega_{\times n-k}) = \frac{1}{Z_n}\sum_{x^{(n)} \in A_k\times \Omega_{\times n-k}} \mathrm{e}^{H(x^{(n)})} \nonumber\\
	&= \frac{1}{Z_n} (\sum_{x^{(k)} \in A_k}\mathrm{e}^{H(x^{(k)})})(2\cosh J)^{n-k}= \frac{1}{Z_k}\sum_{x^{(k)} \in A_k}\mathrm{e}^{H(x^{(k)})} = \nu_H^{(k)} ( A_k)\,.   
\end{align} 
Therefore, the system of measure spaces $\left\lbrace (\Omega_n, \nu_H^{(n)}, \pi_{n,k}) \right\rbrace_{n,k}$ forms an inverse system of compatible probability spaces, and by Kolmogorov theorem~\cite{kolmogorov2013foundations,Shiryaev1996,Rao1971} there is a unique probability measure $\nu_H$ on the set $\Omega_{\infty}$ which can be computed on the cylinder $\mathscr{C}(A_k)$ as follows:
\begin{equation}
	\nu_H(\mathscr{C}(A_k)) = \nu_H^{(k)}(A_k). 
\end{equation}
Moreover, this measure $\nu_{H}$ satisfy the following covariance property under the action of the group $\Gamma$:
\begin{align}
	((L_{x^o})_{\ast}\nu_H)(\mathscr C((y_1,\dots,y_n))) &=\nu_H (\mathscr C(y_1\oplus x ^o_1,\dots,y_n\oplus x^o_n))\nonumber\\
	& = \mathrm{e}^{H(y_1\oplus x ^o_1,\dots,y_n\oplus x^o_n)} = \Delta_H^{-1}(x^o,y) \mathrm{e}^{H(y_1,\dots,y_n)}\,
\end{align}
so that $((L_{x^o})_{\ast}\nu_H) = \Delta_H^{-1} \nu_H$, with $\Delta_H$ given by equation~\eqref{eq:modular_function_easing}.
Summarizing the previous results, we have that the measure $\mu_H = \nu_H\times \#$ defines a left-invariant system of Haar measures for the infinite qubit-chain groupoid $\mathcal{G}$ with modular function $\Delta_H$.  

The multiplication operator by the function $\Delta_H$ defines a self-adjoint operator on the Hilbert space $\mathrm{L}^2(\mathcal{G},\mu_H)$, which is the Tomita-Takesaki operator associated with the new left-Hilbert algebra $\mathfrak{C}$ of the groupoid\footnote{This left-Hilbert algebra is constructed as in Sec.~\ref{ch:infinite_qubit_chain_groupoid} by replacing the Hilbert space $\mathrm{L}^2(\mathcal{G},\mu)$ with the Hilbert space $\mathrm{L}^2(\mathcal{G},\mu_H)$. With a slight abuse of notation we still call it $\mathfrak{C}$.}. We can write explicitly the unitary evolution corresponding to the Hamiltonian $S$:
\begin{equation}
	U_t =\mathrm e^{\mathrm i S t}.
\end{equation}

The corresponding Tomita-Takesaki dynamics~\cite{TakesakiII} can be written for the elements in the left-Hilbert algebra $\mathfrak{C}$ of~\eqref{eq:left_hilbert_algebra_qubit_groupoid}. Let $F$ be in $\mathfrak C$, and write its evolution at time $t$, applied to a function $\psi$ in $\mathrm L^2(\mathcal G, \mu_H)$:
\begin{align}
	\mathrm e^{-\mathrm i S t}F\star\mathrm  e^{\mathrm i S t}\psi(x,x^o)&=\mathrm e^{-\mathrm i S(x,x^o)t}\sum_{y^o\in\Gamma} F(x,y^o)\mathrm  e^{\mathrm i S(x\oplus y^o,x^o\oplus y^o)t}\psi(x\oplus y^o, x^o\oplus y^o)\nonumber\\
	&=\sum_{y^o\in\Gamma} F(x,y^o)\mathrm e^{ -\mathrm i S(x,x^o)t+\mathrm i S(x,x^o)t+\mathrm i S(x,y^o)t} \psi(x\oplus y^o, x^o\oplus y^o)\nonumber\\
	&=	(\mathrm e^{\mathrm i S t}F)\star\psi(x,x^o).
\end{align}

We obtain the interesting result that the Tomita-Takesaki evolution corresponding to the Hamiltonian~\eqref{eq:differential_hamiltonian} is equivalent to the unitary Heisenberg evolution of the elements in the algebra. This property is a direct consequence of the DFS composition~\eqref{eq:dfscompositioncondition_generalformula}, and holds for all DFS function $\mathscr S$. Clearly, the evolved state $\mathrm e^{\mathrm i S t}F$ is still in $\mathfrak C$, since the multiplication operator has modulus $1$, and does not affect neither the $\mathrm L^2$ nor the $\mathrm L^1_\mathrm H$ norms~\eqref{eq:hanhnorminfinitequbitgroupoid}. This is a feature deriving from the fact that by using the measure on the groupoid one defines a left-Hilbert algebra which has a canonical normal weight and an associated modular automorphism.

The previous discussion suggests, in a simple model, that the DFS functions are central in the description of dynamics in the groupoid picture not only for the groupoid of histories. Indeed, we have seen that these functions are associated with the unitary evolution determined by the Tomita-Takesaki operator of the left-Hilbert algebra used for the definition of the groupoid algebra. This evolution is naturally associated with the choice of a left-invariant system of Haar measures on the groupoid of interest. In this sense, it is a thermodynamical time for the system under analysis~\cite{Connes1994}.  

\section{Conclusions}
In this paper we analyzed some features of the groupoid picture of quantum mechanics. In particular, we focused on the construction of algebras of observables which would be non-trivial. Using the action-groupoid built out of a free, ergodic and non-transitive action of a discrete countable group $\Gamma$ on a measurable space $\Omega_{\infty}$, we explicitly showed that the associated von Neumann reduced algebra is a factor which could be type $II_1$ or type $III_{\lambda}$ depending on the choice of the measure on $\Omega_{\infty}$. This construction reproduces well-known results from the theory of von Neumann algebras and in this work we have tried to interpret them from the perspective of Schwinger's picture of quantum mechanics. In this direction, we have provided a different interpretation for the positive definite function $\varphi_\mathscr S = \exp{\mathrm i\mathscr S}$ which are associated with the so-called DFS states. We have seen that replacing the imaginary exponential with a real exponential we obtained a homomorphism of the groupoid with values in the the group of positive real numbers and we can find a measure for which this is the corresponding modular function. Therefore, this modular function generates a modular automorphism which determines a canonical dynamics associated with the choice of the measure on the groupoid. The parameter of this dynamics is a thermal time for the system under investigation. As a particular example of this construction we have built a measure on the groupoid $\mathcal{G}=\Omega_{\infty}\times\Gamma$ whose associated modular automorphism can be interpreted as the dynamics of an infinite chain of spins with an Ising interaction. In a future work we would like to interpret this groupoid as emerging from the action of a ``gauge'' group on the groupoid of histories of a simplified configuration space made up only of two outcomes. In this sense, we could interpret the DFS function as an action functional on this configuration space. This implementation could help to clarify some aspects of Schwinger's picture of quantum mechanics: from one point of view it could shed some light on the formulation of the quantum action principle, from the other point of view it could be a toy model of a system subject to local gauge transformations. 

\section*{Acknowledgments}
This work was partially supported by Istituto Nazionale di Fisica Nucleare (INFN) through the project ``QUANTUM'' and the Italian National Group of Mathematical Physics (GNFM-INdAM).  The authors acknowledge financial support from the Spanish Ministry of Economy and Competitiveness, through the Severo Ochoa Programme for Centres of Excellence in RD (SEV-2015/0554), the MINECO research project  PID2020-117477GB-I00,  and Comunidad de Madrid project QUITEMAD++, S2018/TCS-A4342.
GM would like to thank partial financial support provided by the Santander/UC3M Excellence  Chair Program 2019/2020, and he is also a member of the Gruppo Nazionale di Fisica Matematica (INDAM), Italy. 
FDC thanks the UC3M, the European Commission through the Marie Sklodowska-Curie COFUND Action (H2020-MSCA-COFUND-2017- GA 801538) and Banco Santander for their financial support through the CONEX-Plus Programme. 
FMC acknowledges that the work has been supported by the Madrid Government (Comunidad de Madrid-Spain) under the Multiannual Agreement with UC3M in the line of ``Research Funds for Beatriz Galindo Fellowships'' (C$\setminus$\&QIG-BG-CM-UC3M), and in the context of the V PRICIT (Regional Programme of Research and Technological Innovation).

\bibliography{mybibfile.bib}
\bibliographystyle{ieeetr}

\end{document}